\documentclass[aps,pra,superscriptaddress,reprint, twocolumn, amsmath, amssymb, a4, notitlepage,longbibliography]{revtex4-1}
\usepackage{graphicx}
\usepackage{amsmath,amssymb}
\usepackage{dcolumn}
\usepackage{mathrsfs}
\usepackage{physics}
\usepackage{float}
\usepackage{booktabs}
\usepackage{svg}
\usepackage{comment}
\usepackage{color}
\usepackage{tikz}
\usepackage{braket}
\usepackage{pgfplots}
\usepackage{standalone}
\usetikzlibrary{positioning,fit}

\usepackage[dvipsnames]{xcolor}

\usepackage[colorlinks]{hyperref}

\hypersetup{
    colorlinks=true,
    linkcolor=blue,
    filecolor=magenta,      
    urlcolor=magenta,
    citecolor={blue},
    }

\usetikzlibrary{calc}

\begin{document}
\title{Decoherence in Waveguide Quantum Electrodynamics using Matrix Product States}
\date{\today}
	
\author{Matias Bundgaard-Nielsen}
\email[]{matbu@dtu.dk}
\affiliation{Department of Electrical and Photonics Engineering, Technical University of Denmark, Building 343, 2800 Kongens Lyngby, Denmark}
\affiliation{NanoPhoton - Center for Nanophotonics, Technical University of Denmark, Building 343, 2800 Kongens Lyngby, Denmark}
\author{Matthew Kozma}
\affiliation{Department of Physics, Engineering Physics and Astronomy, Queen’s University, Kingston ON K7L 3N6, Canada}
\author{Sofia Arranz Regidor}
\affiliation{Department of Physics, Engineering Physics and Astronomy, Queen’s University, Kingston ON K7L 3N6, Canada}
\author{Stephen Hughes}
\affiliation{Department of Physics, Engineering Physics and Astronomy, Queen’s University, Kingston ON K7L 3N6, Canada}

\begin{abstract}
We present a matrix product state (MPS) method for including decoherence processes in calculations involving waveguide quantum electrodynamics (waveguide QED) using density matrices. The approach is based on collision quantum optics, where the many-body state of the waveguide is represented using discrete time bins, which are then efficiently encoded using an MPS chain. Our method is a generalization of previous MPS methods, and we demonstrate how one can efficiently extend the method to density matrices, allowing for the inclusion of various loss processes in the form of Lindblad terms in the Liouvillian superoperator responsible for the relevant dissipation dynamics. As an application of the theory, we study various waveguide QED systems and the influence of emitter pure dephasing (which is one of the most important processes in real systems) on the light-matter interactions, including a two-level system (TLS) in a semi-infinite waveguide with time-delayed feedback, two spatially separated TLSs with finite delays, and finally the scattering of few-photon Fock pulses on a TLS. In addition to emitter pure dephasing, we also show how to include off-chip radiative decay, and show how it differs qualitatively from pure dephasing.   
 
\end{abstract}
\maketitle

\section{Introduction}
Waveguide quantum electrodynamics (waveguide QED) provides a versatile platform for studying light–matter interactions with quantum circuits and waveguides, where localized quantum emitters couple to a continuum of propagating field modes. Such systems enable strong and highly controllable photon-mediated interactions and have been explored in both nanophotonic systems \cite{Coles2016,Trschmann2018,Dalacu2019,LeJeannic2021ExperimentalWaveguide,LeJeannic2022DynamicalEmitter,Siampour2023,Patel2026} and superconducting circuit architectures
\cite{Hoi2015,Mirhosseini2019,Ferreira2024,Grebel2024BidirectionalNodes,Cheng2025,Odeh2025,2603.28004}. Their ability to engineer long-range interactions, chiral couplings, and (non-Markovian) time-delayed feedback has made waveguide QED a promising setting for quantum information processing, quantum networking, and the generation of nonclassical light \cite{Pichler2016PhotonicFeedback,Pichler2017UniversalFeedback,Ferreira2024}.

The theoretical description of waveguide QED systems is inherently challenging, especially beyond the linear or weak-excitation regime. Even in linear regimes where the emitter response can often be described in terms of single-excitation scattering or effective susceptibilities, waveguide-QED experiments have revealed a range of interference, collective, and retardation effects, including mirror-induced modification of radiative decay  \cite{Hoi2015}, photon-mediated interactions \cite{vanLoo2013Photon-MediatedAtoms}, collective Lamb shifts between spatially separated superconducting qubits \cite{Wen2019LargeAtoms}, and time-delayed coherent feedback leading to genuinely non-Markovian dynamics \cite{Ferreira2021CollapseReservoir,2603.28004}. These experiments already require theoretical descriptions that account for finite propagation times, interference, and structured electromagnetic environments. The challenge becomes even more pronounced in nonlinear few-photon and strongly driven regimes, where the two-level nature of the emitter cannot be linearized, and the scattered field can develop strong photon--photon correlations. Such effects are central to recent experiments on few-photon scattering from quantum dots \cite{LeJeannic2021ExperimentalWaveguide,LeJeannic2022DynamicalEmitter,Patel2026}, as well as experiments on superconducting artificial atoms where time-delayed coherent feedback modifies nonlinear resonance fluorescence and Mollow-triplet spectra \cite{2603.28004}. They also play an important role in multi-photon state generation using waveguide-mediated feedback \cite{Ferreira2024}.

From a theoretical point of view, the difficulty is that the electromagnetic field forms a continuum of modes, and the joint emitter–field state rapidly grows in the size of the Hilbert space. In many situations, the Markov approximation provides a useful simplification, leading to simplified master equations with instantaneous couplings \cite{Caneva2015QuantumFormalism,Combes2017TheNetworks, Arranz2023Probingpumping}. However, this approximation neglects finite propagation times and memory effects, which can become important when emitters are spatially separated or when (delayed) feedback is present. In such regimes, non-Markovian dynamics, time-retardation effects, and strong photon–photon correlations can play a central role, and are essential to include in the modeling.

Recently, tensor-network methods, especially matrix product state (MPS) approaches, have emerged as powerful tools for simulating non-Markovian waveguide QED dynamics  \cite{Grimsmo2015Time-DelayedControl,Pichler2016PhotonicFeedback,Pichler2017UniversalFeedback,ArranzRegidor2021ModelingModel,ArranzRegidor2021CavitylikeRegime,PhysRevResearch.3.023168,Vodenkova2024,ArranzRegidor2025TheoryExcitations,ArranzRegidor2025TheorySpectra}. By discretizing the field into {\it time bins} and exploiting the limited entanglement structure of one-dimensional systems, MPS techniques 
enable {\it numerically exact simulations} of few-emitter systems with delayed feedback and multiple excitations.

Most existing MPS-based treatments of waveguide QED, however, focus on 
{\it pure-state dynamics} of the joint emitter–field system. In these approaches, the evolved state remains pure, and losses are generally not included. While photon losses can be incorporated through the introduction of additional ancilla waveguide channels, pure dephasing of the emitter cannot be included directly, and pure dephasing is one of the most experimentally relevant processes to include, for semiconductor systems as well as circuit-QED~\cite{2603.28004,Hoi2015,Mirhosseini2019,Ferreira2024,Cheng2025,Odeh2025}. Although examples of implementations with the possibility of losses exist \cite{PhysRevResearch.3.023168,Vodenkova2024}, the numerical complexities, limitations of the implementations, and novelty of the methods have prevented studies of the impact of pure dephasing, especially when the input state of the waveguide is not vacuum.  

An alternative strategy is to average over many stochastic trajectories, but this quickly becomes numerically costly and introduces additional overhead in managing multiple trajectories. Stochastic simulation with quantum trajectories and a discrete waveguide model is
powerful when limiting the number of waveguide quanta~\cite{ArranzRegidor2021ModelingModel,Whalen2019,PhysRevA.106.013714,PhysRevA.110.L031703,2603.28004}, but is not well suited for many-body waveguide QED (many quanta regime).

Thus, while such approaches are appropriate for near-ideal waveguide interfaces, realistic experimental platforms inevitably involve additional decoherence processes. Accurately capturing the interplay between decoherence and non-Markovian waveguide dynamics, therefore, remains largely unexplored.

An efficient representation of the 
waveguide QED density matrix is thus a natural route for including decoherence. One possibility is to use locally purified density operators (LPDO), where positivity of the density matrix is preserved by construction through additional local Kraus or ancilla indices \cite{Verstraete2004,Werner2016,Cheng2021,Guo2024,Guo2024Nature}. This can be advantageous in simulations of large noisy quantum circuits. However, the additional auxiliary dimensions also make the implementation more involved, and recent work has shown that LPDO simulations can become numerically challenging in regimes where noise and operator-space entanglement compete strongly \cite{Guo2024}.

In this work, we extend the existing time-bin waveguide QED MPS frameworks to use a vectorized density matrix MPS representation of the waveguide QED system. This generalization allows us to simulate non-Markovian waveguide dynamics with the inclusion of photon loss, pure dephasing, incoherent pumping, and other dissipative processes implemented through Lindblad decoherence terms. Our approach keeps an efficient MPS representation of the emitter-waveguide state that allows for manageable computation times, and immediately renders the MPS approach much more powerful, as all experimental systems have such losses. Moreover, since photons are treated exactly, the role of pure dephasing in waveguide QED is not a simple decay process, enabling us to explore fundamentally new regimes of light-matter interaction with real dephasing mechanisms included. Importantly, this also provides access to the {\it operator entanglement}, which directly relates to the classical simulability of the dynamics \cite{Dubail2017Entanglement1+1d,Wellnitz2022RiseDephasing}. In this sense, our framework enables an assessment of whether proposed photonic protocols can reach regimes of computational complexity beyond efficient classical simulation.

In particular, we focus on the impact of pure dephasing in two-level systems (TLSs) and how it affects time-delayed feedback dynamics as well as the scattered fields themselves. We consider three important examples. First, a TLS in a semi-infinite waveguide experiencing time-delayed feedback. This system is known to exhibit {\it perfect} population trapping when the delayed field has the appropriate phase. We show that pure dephasing disrupts this destructive interference, leading to a slow decay of the otherwise trapped population. For completeness, we also show that the off-chip decay can be easily included, with effects that are more obvious. 

Second, we consider two spatially separated emitters, which exhibit non-Markovian waveguide-mediated superradiance. Here, pure dephasing unsurprisingly suppresses the coherent emitter–emitter interaction and thereby reduces the superradiant behavior. We further analyze the buildup of quantum correlations between the emitters using entanglement measures, and discuss the role of operator entanglement as a measure of the complexity of the joint emitter–field state.

Finally, we study the scattering of one- and two-photon Fock-state pulses from a single chirally coupled two-level system. In this case, pure dephasing leads to nontrivial modifications of the scattered field, including qualitative changes to the emitted spectrum.

The rest of the paper is organized as follows. In Sec.~\ref{sec:timebin}, we introduce the model and describe the density-matrix MPS theoretical framework, including the time-bin representation, the MPS decomposition, and the time-evolution scheme for both Markovian and non-Markovian dynamics. In Sec.~\ref{sec:tls_nmark}, we study a single emitter subject to time-delayed feedback and analyze the effect of pure dephasing on population trapping. In Sec.~\ref{sec:2tls}, we consider two emitters coupled through the waveguide, investigating both superradiant dynamics and the emergence of entanglement and operator entanglement. In Sec.~\ref{sec:fockpulses}, we turn to the scattering of few-photon Fock pulses from a chirally coupled emitter and analyze how pure dephasing modifies the transmitted field and its spectrum. Finally, we conclude in Sec.~\ref{sec:conclusion}. In Appendix \ref{app:dm_mps_pulses}, we discuss the density matrix representation of few-photon Fock states. In Appendix \ref{app:numerical_details}, we provide numerical details on the MPS simulations. In Appendix \ref{app:chiral}, we derive an analytical expression for the stationary emission spectrum of a single-photon scattered on a chiral emitter under the influence of pure dephasing.

\section{Model and Theory}
In this section, we introduce the general model of a localized system interacting with the continuum of the waveguide and show how, under certain conditions, we can represent the waveguide state using time bins. This is sometimes also referred to as collision quantum optics \cite{Ciccarello2018CollisionOptics,Whalen2019}. We then introduce MPS with density matrices as an efficient way of representing these time bin states.

\begin{figure}[!ht]
    \centering
    \includegraphics[width=\linewidth]{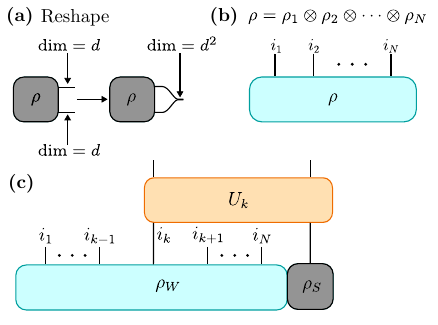}
    \caption{(a) A density matrix of dimensions $(d,d)$ is reshaped to a vector of length $d^2$. (b) The tensor product of $N$ density matrices is represented as a rank-N tensor [with flattened legs as shown in (a)]. (c) The time evolution of the waveguide and system is calculated by applying $U_k$ to the system and $k$'th waveguide bin. }
    \label{fig:bins_sketch}
\end{figure}

\subsection{Time-Bin Picture of Waveguide QED \label{sec:timebin}}
We consider a single quantum emitter (TLS) interacting with a left and right propagating waveguide mode in the following derivation, but the derivation is easily extended to multiple emitters and potentially more modes. The Hamiltonian reads:
\begin{equation}
    H = H_S + H_W + H_{SW},    
\end{equation}
where (we work in natural units with $\hbar=1$) $H_S =  \omega_0 \sigma ^\dagger \sigma$ is the system Hamiltonian of the TLS with $\sigma = \ket{g}\bra{e}$ being the transition operator; $H_W = \sum_{\alpha=L,R} \int d\omega \  \omega \ b_\alpha^\dagger(\omega) b_\alpha(\omega)$, is the free energy of the two waveguide modes with $b_\alpha(\omega)$ being the annihilation operator of a photon with frequency $\omega$ in waveguide mode $\alpha$ (either left or right), obeying $\comm{b_\alpha(\omega)}{b_\beta^\dagger(\omega')} = \delta_{\alpha\beta}\delta(\omega - \omega')$; finally, the interaction between the emitter and waveguide is 
\cite{Ciccarello2018CollisionOptics}
\begin{equation}
    H_{SW} = \sum_{\alpha=L,R} \int \mathrm{d} \omega  \left(g_\alpha(\omega) \sigma^{\dagger} b_\alpha(\omega) + g_\alpha(\omega)^{*} \sigma b_\alpha^{\dagger}(\omega)\right),
\end{equation}
where $g_\alpha(\omega)$ is the light-matter coupling rate between the emitter and photons of frequency $\omega$ in waveguide $\alpha$. 

We then assume that the coupling $g_\alpha(\omega) = g_\alpha(\omega_0) =\sqrt{\gamma_\alpha/2\pi}$ is frequency independent, which is reasonable when the local density of states of the waveguide is spectrally flat around the emitter frequency $\omega_0$. Moving into the interaction picture with respect to $H_S+H_W$, then \cite{Ciccarello2018CollisionOptics,Bundgaard-Nielsen2024WaveguideQED.jl:Electrodynamics}
\begin{equation}
    H_{SW}(t) = \sum_\alpha \sqrt{\gamma_\alpha} \left ( b_\alpha(t) \sigma^\dagger + b_\alpha^\dagger(t) \sigma \right ), 
\end{equation}
where $H_{SW}(t) = \mathrm{e}^{i( H_S + H_W)t } H_{SW} \mathrm{e}^{-i( H_S + H_W)t } $, and $b_\alpha(t) = \int d\omega  \ b_\alpha(\omega) \mathrm{e}^{i(\omega - \omega_0) t }$, which still obeys $\comm{b_\alpha(t)}{b_\beta^\dagger(t')} = \delta_{\alpha\beta} \delta(t - t')$. We now see that we have gone from an interaction between the emitter and a continuum of frequency modes to an interaction between time-dependent field modes $b_\alpha(t)$ and the emitter. This structure can be exploited using matrix product states, as we will see in the next section, since the interaction is now localized to single (time-bin) modes at a time, enabling an efficient representation of both the waveguide and the system evolution.  

Before we introduce the MPS method, however, we need to discretize the time bin modes. We start by considering the time evolution of the density matrix,
\begin{equation}
    \frac{d \rho}{d t} = -i[H(t),\rho] + \mathcal{D}[\rho] = \mathcal{L}(t)[\rho] ,\label{eq:rho_eom}
\end{equation}
where $\mathcal{D}[\rho]$ represents any decoherence processes in the system, which can be modeled through Lindblad terms. We then consider the time-evolution operator $U(t_0,t)$ which evolves the state $\rho(t) = U(t_0,t) [\rho(t_0)]$ from time $t_0$ to $t$, which is formally given by \cite{Breuer2007TheSystems}: 
\begin{equation}
    U(t_0,t)[\cdot] = \mathcal{T} \exp \left \{\int_{t_0}^t dt' \mathcal{L}(t') [\cdot] \right \},  
\end{equation}
where $\mathcal{T}$ is the time ordering operator and $[\cdot]$ represents whatever object $U(t_0,t)$ is applied to. We then discretize the time into $N$ bins with $t = N\Delta t $ and $t_k = k \Delta t$ and using the Suzuki-Trotter decomposition we can then write: $U(t_0,t) = \lim \limits_{N\rightarrow\infty} U_N \cdots U_1$ with $U_k = \exp \left \{ \int_{t_{k-1}}^{t_k} dt' \mathcal{L}(t')[\cdot]  \right\}$. 

We subsequently introduce the time-discretized waveguide operator,
\begin{equation}
    b_{\alpha,k} = \frac{1}{\sqrt{\Delta t}} \int_{t_{k-1}}^{t_{k}} d t^{\prime} b_\alpha(t^{\prime}), \label{eq:bak}
\end{equation}
where the factor of $1/\sqrt{\Delta t}$ ensures $\comm{b_{\alpha,k}}{b_{\beta,k'}^\dagger} = \delta_{\alpha\beta} \delta_{k,k'}$, which simplifies the evolution operator to
\begin{equation}
    U_k = \exp \left \{ \Delta t (-i\comm{H_k}{\cdot} + \mathcal{D}[\cdot])  \right\} , \label{eq:disc_U}
\end{equation}
where we defined the effective discretized waveguide Hamiltonian:
\begin{equation}
    H_{k} = \sum_\alpha \sqrt{\gamma_\alpha / \Delta t} \left ( b_{\alpha,k} \sigma^\dagger + b_{\alpha,k}^\dagger \sigma \right ) .
\end{equation}

Thus, we see that, within the precision of the Trotter decomposition, we can reduce the waveguide problem to a sequence of interactions between discretized modes with operators $b_{\alpha,k}$. This allows us to write the state of the entire waveguide and emitter as
\begin{equation}
    \rho = \rho_1 \otimes \rho_2 \otimes\cdots \otimes \rho_k \otimes\cdots \otimes \rho_N \otimes \rho_S,
\end{equation}
where $\rho_k$ denotes the state of the $k$'th waveguide bin containing both the left and right propagating modes. 

In Fig.~\ref{fig:bins_sketch}(a), we show the tensor representation of a density matrix $\rho$, where the two matrix legs are fused into a single index, which is equivalent to reshaping (or flattening) the matrix. In Fig.~\ref{fig:bins_sketch}(b), we illustrate how a density matrix composed of $N$ tensor products can be represented as an $N$-rank tensor (again all density matrices have been flattened according to Fig.~\ref{fig:bins_sketch}(a)). In Fig.~\ref{fig:bins_sketch}(c), we show the total state of the waveguide and how the time-evolution operator in Eq.~\ref{eq:disc_U} is applied at timestep $k$. Note that $U_k$ is here a $d_S^2 d_W^2 \times d_S^2 d_W^2$ matrix reshaped into a rank-four tensor with dimensions $(d_S^2,d_W^2,d_S^2,d_W^2)$, where $d_S$ is the system dimension and $d_W$ is the dimension of the waveguide bins. The squares arise from the flattening of the density matrices.

Although the above formulation is conceptually straightforward, it is often not practical to store the full density matrix of the combined waveguide–system state due to the exponential scaling of the tensor product. One can attempt to construct efficient bases and implement the evolution operator $U_k$ efficiently, which has been done for ket states \cite{Bundgaard-Nielsen2024WaveguideQED.jl:Electrodynamics}. However, even with such techniques, representing the full density matrix quickly becomes infeasible. Instead, we can exploit the fact that matrix product states can efficiently represent high-rank tensors. In the next section, we therefore introduce how the system–waveguide density matrix can be represented in this form and efficiently evolved according to $U_k$.

\subsection{Singular Value Decomposition of Quantum States}

The discretization described above naturally leads to a tensor representation of the joint system–waveguide density matrix, where each site corresponds to either the emitter or a waveguide time bin. However, storing the full tensor quickly becomes infeasible, since the dimension of the Hilbert space grows exponentially with the number of time bins. Fortunately, the sequential interaction structure of the collision model strongly constrains the correlations that can build up along the time-bin chain. This structure can be efficiently captured using matrix product states, which provide a compressed representation of high-rank tensors by factorizing them into a sequence of lower-rank tensors connected by bond indices.

\begin{figure}[!ht]
    \centering
    \includegraphics[width=\linewidth]{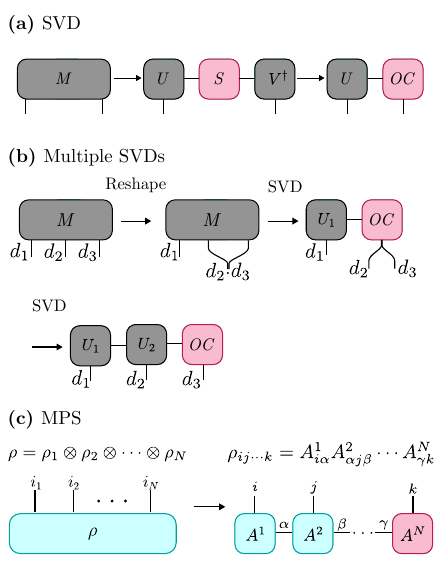}
    \caption{(a) Singular value decomposition in tensor diagrams. We contract the Schmidt coefficients of $S$ into $V^\dagger$, which we denote as the Orthogonality Center (OC) and highlight with a red box. (b) We show the principle of how multiple SVDs are performed to split a rank-3 tensor into its constituent MPS. (c) We show how a density matrix made up of many subsystems can be efficiently represented using MPS.}
    \label{fig:svd_mps}
\end{figure}

The main idea in MPS is to factorize the tensor using {\it singular value decompositions} (SVDs). Mathematically, the SVD can be written as
\begin{equation}
    M = U \cdot S \cdot V^\dagger, \label{eq:SVD}
\end{equation}
where if $M$ is a matrix of dimensions $m \times n$, then $U$ has dimensions $m \times \chi$, $S$ is a diagonal matrix of dimensions $\chi \times \chi$, and $V$ has dimensions $n \times \chi$, with $\chi \leq \min(m,n)$. In Fig.~\ref{fig:svd_mps}(a), we illustrate the singular value decomposition using tensor diagrams. Often, the singular values contained in $S$ are absorbed into one of the tensors $U$ or $V$. The resulting tensor is then referred to as the {\it Orthogonality Center} (OC), which we denote schematically with a red (lighter colored) box.

For a density matrix composed of two subsystems with flattened dimensions $m$ and $n$, this corresponds to an $m \times n$ tensor [see Fig.~\ref{fig:bins_sketch}(b)], which can be decomposed as \cite{Verstraete2004,houdayer2025tensormixedstatesjulialibrarysimulating}:
\begin{equation} 
    \rho = \sum^{\chi}_\alpha 
 \lambda_\alpha \rho^1_\alpha \otimes \rho^2_\alpha ,\label{eq:svd_dm}
\end{equation}
where $\lambda_\alpha$ is the $\alpha$'th diagonal element of $S$ in the SVD, $\rho^1_\alpha$ is the $\alpha$'th column of $U$, and $\rho^2_\alpha$ is the $\alpha$'th column of $V$. If no correlations are present between the subsystems, then $\chi = 1$ and the state reduces to a simple product state. 

We will refer to $\chi$ as the ``bond dimension''. In many quantum systems with structured correlations, $\chi$ can remain much smaller than $m$ and $n$ allowing for an efficient representation, thus motivating the use of the SVD. If the density matrix is composed of multiple subsystems (such as the discretized waveguide state), the SVD can be applied sequentially as illustrated in Fig.~\ref{fig:svd_mps}(b). In this way, the full tensor can be written in the matrix product form,
\begin{equation}
    \rho_{ij\cdots k} = A^1_{i\alpha} A^2_{\alpha j \beta} \cdots A^N_{\gamma k},
\end{equation}
where the subscripts in $\rho_{ij\cdots k}$ denote the indices of the individual subsystems of $\rho$, with $i$, $j$, and $k$ labeling the local flattened density matrix of the corresponding subsystem. 

We use Einstein summation notation, so repeated indices ($\alpha$, $\beta$, $\gamma$) are implicitly summed over. The tensors $A^1_{i\alpha}$ and $A^N_{\gamma k}$ correspond to the first and last sites of the MPS chain, representing the decomposition of the first and last subsystems. Consequently, they do not carry bond indices extending to the left and right, respectively. The intermediate tensors, such as $A^2_{\alpha j \beta}$, connect neighboring sites of the chain and therefore carry two bond indices ($\alpha$ and $\beta$) in addition to the local subsystem index $j$. These bond indices encode the correlations shared between adjacent subsystems. A schematic illustration of this decomposition is shown in Fig.~\ref{fig:svd_mps}(c), where the OC is located in the tensor $A^N$.

In practice, the initial waveguide state is typically known directly in MPS form, and therefore, the sequential SVD procedure described above is not explicitly performed. It is introduced here primarily to illustrate how the MPS representation arises from the structure of the full tensor. For example, if the waveguide Hilbert space is truncated to a maximum of one photon, the vacuum density matrix has a trivial MPS representation with bond dimension $\chi = 1$, corresponding to all sites being in the vacuum state: $A^{i}_{\alpha,:,\beta} = [1,0,0,0]$. In Sec.~\ref{sec:fockpulses}, we discuss how one- and two-photon pulses can be incorporated within this framework.

\begin{figure}[!ht]
    \centering
    \includegraphics[width=\linewidth]{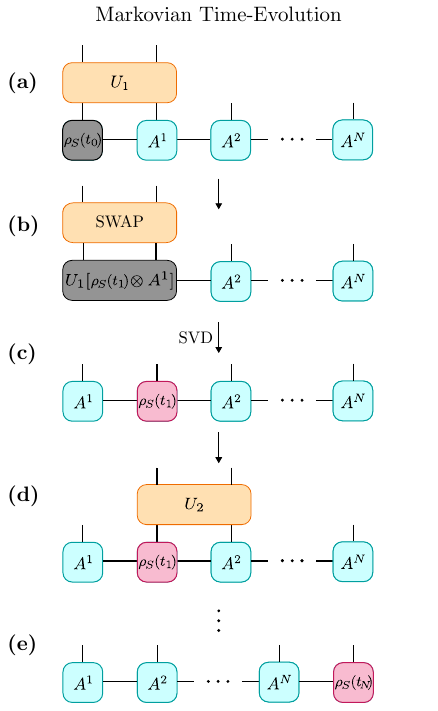}
    \caption{Tensor diagrams showing how the Markovian time evolution is handled. (a) The evolution operator $U_1$ is applied to the system bin $\rho(t_0)$ and the first waveguide bin $A^1$. (b) and (c), the system and waveguide bins are swapped and subsequently decomposed using SVD, bringing the system bin adjacent to the next waveguide bin $A^2$. (d) The next timestep begins with the evolution operator $U_2$ applied to the system bin and the next waveguide bin $A^2$. The rest repeats from (b).}
    \label{fig:mark_evo}
\end{figure}

\subsection{Markovian Time-evolution of the Waveguide}

We now describe how the time evolution of the system–waveguide state is carried out within the MPS representation. As derived in Sec.~\ref {sec:timebin}, the dynamics can be written as a sequence of local evolution operators $U_k$ acting on the system and the $k$'th waveguide time bin. This structure naturally fits the MPS representation since the evolution only involves two neighboring sites at a time.

The algorithm is illustrated in Fig.~\ref{fig:mark_evo}. At the initial timestep, the evolution operator $U_1$ is applied to the system density matrix $\rho_S(t_0)$ and the first waveguide bin $A^1$, as shown in Fig.~\ref{fig:mark_evo}(a). In the tensor network representation, this corresponds to contracting the rank-four tensor $U_1$ with the system tensor and the first waveguide tensor, resulting in a combined tensor containing both degrees of freedom. 

After this contraction, the system tensor is swapped with the waveguide tensor, as illustrated in Fig.~\ref{fig:mark_evo}(b). This step ensures that the system site is moved along the MPS chain so that it becomes adjacent to the next waveguide bin. The resulting tensor is then decomposed using a singular value decomposition, as shown in Fig.~\ref{fig:mark_evo}(c). The singular values are returned in descending order, allowing truncation of the bond dimension by retaining the largest $\chi$ values while discarding the smallest ones, ensuring an accurate decomposition. At each truncation step, we keep the bond dimension $\chi$ such that
\begin{equation}
    \frac{\sum_{i=1}^{\chi} s_i}{\sum_j s_j} \geq 1-\epsilon_{\mathrm{rel}},
\end{equation}
where $\epsilon_{\mathrm{rel}}$ is the relative cutoff, meaning that the discarded singular values contribute at most a fraction $\epsilon_{\mathrm{rel}}$ of the total sum of singular values. We furthermore impose the additional constraint that $\chi \leq \chi_{\max}$ as we have found that this leads to more numerical stability. As also noted in the introduction, this truncation does not guarantee positivity, and choosing $\chi_{\max}$ too small may therefore lead to unphysical results. In practice, we choose $\epsilon \approx 10^{-7}$ and $\chi_{\max}$ large enough so that it does not affect the dynamics, i.e., further increasing $\chi_{\max}$ produces no visible changes in the results. 
Furthermore, we monitor the Hermiticity error defined as:
\begin{equation}
    \epsilon_{\mathrm{Herm}} =
\frac{\left\lVert \rho - \rho^\dagger \right\rVert_F}
{\left\lVert \rho \right\rVert_F},
\end{equation}
where \(\lVert A \rVert_F = \sqrt{\operatorname{Tr}(A^\dagger A)}\) is the Frobenius norm, and the trace-preservation error, defined as
\begin{equation}
    \epsilon_{\mathrm{tr}} =
\left| \operatorname{Tr}(\rho) - 1 \right|.
\end{equation}
For most simulations, we observe that when $\max(\epsilon_\mathrm{Herm},\epsilon_\mathrm{tr})< 5 \cdot 10^{-3}$, no visible changes occur and the simulations seem to be converged.

Following the decomposition, the system tensor is now positioned next to the second waveguide bin $A^2$. The next timestep, therefore, begins by applying the operator $U_2$ to the system tensor and $A^2$, as illustrated in Fig.~\ref{fig:mark_evo}(d). The same sequence of tensor contractions, swaps, and SVD decompositions is then repeated for each subsequent time bin.

This way, the system tensor effectively propagates along the MPS chain while sequentially interacting with each discretized waveguide mode. Since each timestep only involves local tensor operations and the bond dimension remains limited by the correlations generated during the evolution, the full system–waveguide dynamics can be simulated efficiently even for a large number of time bins and for density matrices.

\subsection{Non-Markovian Time-Evolution}
\begin{figure}[!ht]
    \centering
    \includegraphics[width=\linewidth]{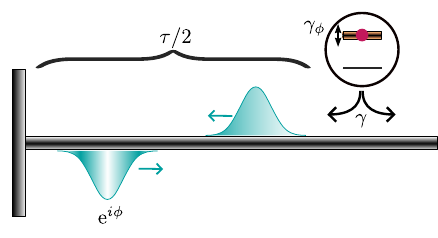}
    \caption{A TLS coupled to a semi-infinite waveguide with coupling rate $\gamma$. A mirror at one end of the waveguide returns emitted photons after a delay of $\tau$. The emitter also experiences a pure dephasing rate of $\gamma_\phi$.}
    \label{fig:sketch_tls_nmark}
\end{figure}

\begin{figure}[!ht]
    \centering
    \includegraphics[width=\linewidth]{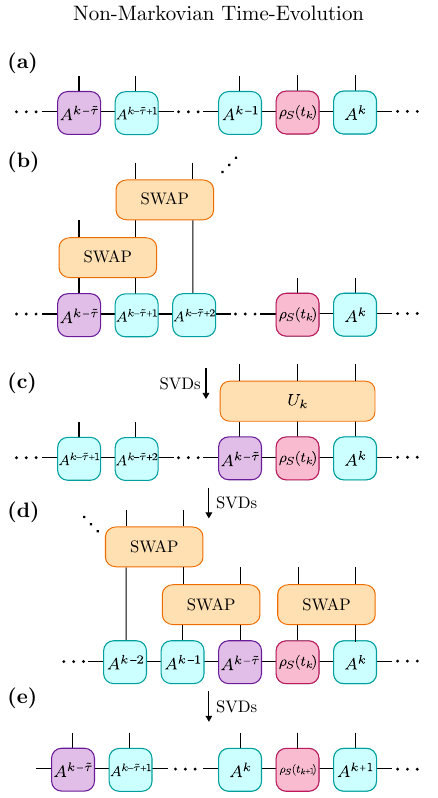}
    \caption{Tensor diagrams showing how the non-Markovian time evolution is performed. (a) We have our system at some time $t_k$ and need to bring the feedback bin (purple) at time $t_k - \tau$ next to the system bin and current waveguide time bin $A^k$. (b) A series of swaps and SVDs brings the feedback bin at index $\tilde \tau  = \tau / \Delta t$ adjacent to the system bin. (c) The time evolution operator $U_k$ is applied to the system bin, feedback bin, and current waveguide bin. (d) After the time evolution and SVD have been performed, the feedback bin is brought back to its original place in the chain. Time evolution then continues from (a) with $k \rightarrow k+1$.
    }
    \label{fig:nmark_evo}
\end{figure}

The time-bin representation introduced above assumes that the emitter interacts only with the present waveguide mode. This corresponds to the Markovian case where excitations emitted into the waveguide do not return to the system. However, there are situations with delayed interactions, giving rise to non-Markovian dynamics, with emissions returning to the emitter.

A canonical example is an emitter coupled to a semi-infinite waveguide terminated by a mirror, with a finite emitter--mirror distance, causing delay-modified dynamics. Photons emitted by the emitter propagate along the waveguide, are reflected by the mirror, and return to interact with the emitter after a propagation time $\tau$ determined by the emitter–mirror distance. See Fig.~\ref{fig:sketch_tls_nmark} for a sketch. In such a setup, the emitter therefore interacts both with the present field and with the field emitted at an earlier time. It can be shown that this leads to an effective interaction between the system and the field operators $b(t)$ and $b(t-\tau)$ \cite{ArranzRegidor2021ModelingModel}. The Hamiltonian used in each timestep thus becomes \cite{ArranzRegidor2021ModelingModel,Bundgaard-Nielsen2024WaveguideQED.jl:Electrodynamics}:
\begin{equation}
\begin{aligned}
    H_{k} &= \sqrt{\frac{\gamma }{2\Delta t }} \left ( b_k \sigma^\dagger + b_k^\dagger \sigma \right )  \\ 
    &+\sqrt{\frac{\gamma}{2 \Delta t}} \left ( \mathrm{e}^{i \phi}b_{k-\tilde \tau} \sigma^\dagger + \mathrm{e}^{-i \phi} b_{k-\tilde \tau}^\dagger \sigma \right ) ,\label{eq:tls_nmar_hamiltonian}
\end{aligned}
\end{equation}
where we here defined the index of the time delay $\tilde \tau = \tau / \Delta t$ (number of discretized time units for propagation) and the phase of the mirror $\phi$. 

Thus, this delayed interaction corresponds to a coupling between the system, the current waveguide bin $A^k$, and the waveguide bin located $\tilde{\tau}$ time steps earlier. Consequently, the evolution operator $U_k$ now acts simultaneously on three sites: the system, the current waveguide bin, and the delayed feedback bin. Note that here we only need one waveguide mode to describe the delayed feedback and have dropped the first subscript $\alpha$ in the waveguide operators $b_{\alpha,k}$ in Eq.~\ref{eq:bak}, denoting left or right.

Implementing this interaction within the MPS representation requires bringing the delayed bin adjacent to the system site before applying the evolution operator. The procedure is illustrated in Fig.~\ref{fig:nmark_evo}. At time $t_k$, the system tensor interacts with the current waveguide bin $A^k$, while the feedback bin located $\tilde{\tau}$ sites earlier contains the field that will reinteract with the system (the delayed bin contains the field emitted at time $t-\tau$).

As shown in Fig.~\ref{fig:nmark_evo}(a), the feedback bin initially resides at position $k-\tilde{\tau}$ in the MPS chain. A sequence of swap operations and intermediate SVD decompositions is therefore performed to move this bin along the chain until it becomes adjacent to the system tensor, as illustrated in Fig.~\ref{fig:nmark_evo}(b). Once the system bin, feedback bin, and current waveguide bin are in neighboring sites, the evolution operator $U_k$ can be applied, as shown in Fig.~\ref{fig:nmark_evo}(c).

After the interaction and subsequent SVD decomposition, the feedback bin is moved back to its original position in the chain through the reverse sequence of swap operations, as illustrated in Fig.~\ref{fig:nmark_evo}(d). The system tensor is then positioned next to the next waveguide bin $A^{k+1}$, and the time evolution proceeds to the next timestep with $k \rightarrow k+1$.

In this way, delayed feedback processes can be incorporated within the MPS framework while preserving the local structure of the time-bin representation. The additional swap operations and the correlations generated by the delayed interaction typically lead to larger bond dimensions as more entanglement is generated, but are still manageable.

\subsection{Expectation Values}

\begin{figure}[!ht]
    \centering
    \includegraphics[width=\linewidth]{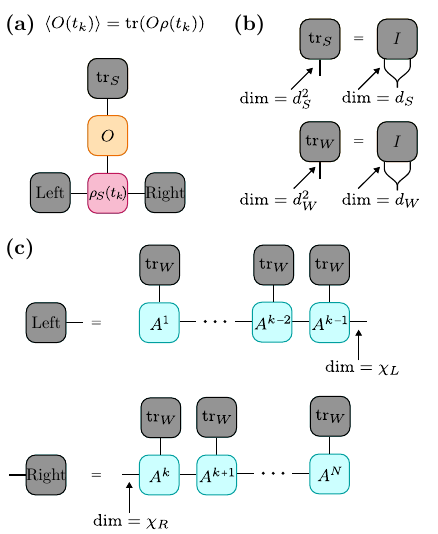}
    \caption{Tensor-network representation of the expectation value of a system operator. 
    (a) Contraction used to compute the expectation value of an operator $O$ acting on the system bin at time $t_k$. The physical index of the system tensor is contracted with $O$, and the resulting tensor is traced using the trace vector $\mathrm{tr}_S$. The bond indices are contracted with the traced environments of the waveguide bins to the left and right. 
    (b) Illustration of the trace vectors used in the contraction, corresponding to flattened identity matrices. 
    (c) Definition of the left and right environments obtained by tracing over the waveguide bins to the left and right of the system site.
     }
    \label{fig:exp_val}
\end{figure}

Once the system–waveguide density matrix has been represented as an MPS, expectation values of observables can be computed efficiently through tensor-network contractions. For an operator $O$ acting on the system bin, the expectation value at time $t_k$ is given by
\begin{equation}
    \langle O(t_k) \rangle = \mathrm{Tr}\!\left[ O \, \rho(t_k) \right].
\end{equation}

Since each density matrix has been flattened, the trace operation can be written as a contraction between the tensor network representing $\rho$ and a trace vector corresponding to the flattened identity matrix. We denote this trace vector by $\mathrm{tr}_S$ for the system and $\mathrm{tr}_W$ for the waveguide.

The contraction used to compute expectation values is illustrated in Fig.~\ref{fig:exp_val}(a). The physical index of the system tensor is first contracted with the operator $O$, after which the trace over the system degrees of freedom is performed by contracting with the trace vector $\mathrm{tr}_S$. The remaining bond indices connect the system tensor to the rest of the MPS chain.

To evaluate the full trace, the waveguide degrees of freedom must also be contracted. This can be done efficiently by defining left and right environments that correspond to tracing over all waveguide bins to the left and right of the system site, respectively. These environments are shown schematically in Fig.~\ref{fig:exp_val}(c).

In practice, the left environment is built iteratively during the time evolution by successively contracting the outgoing waveguide bins. Similarly, the right environments can be constructed once at the beginning by contracting the waveguide bins from the right, and then stored for later use. In this way, both environments are readily available when evaluating observables at any given time.

The evaluation of expectation values then reduces to contracting a small tensor network consisting of the system tensor, the operator $O$, and the corresponding left and right environments, which is minimal in computational cost.

\subsection{Two-time expectation values}

In addition to single-time observables, in quantum optics, one is often interested in two-time expectation values to characterize the out-coupled field, for example, in the emission spectrum, as discussed in Sec.~\ref{sec:fockpulses}. Since we have access to the full system–waveguide state, two-time correlations of the form $\expval{X(t) Y(t+\tau)}$ can be computed directly, without the use of the quantum regression theorem. 

\begin{figure}
    \centering
    \includegraphics[width=\linewidth]{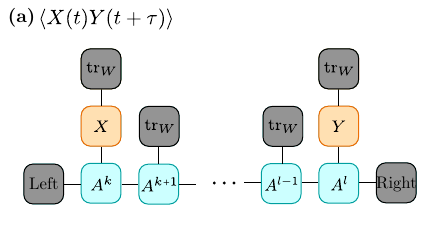}
    \caption{Tensor-network representation of the two-time expectation value $\expval{X(t)Y(t+\tau)}$, with $X$ and $Y$ acting on the waveguide bins and where the left and right tensors are given in Fig.~\ref{fig:exp_val}. The indices $k$ and $l$ correspond to the discretized times $t = k \Delta t$, $t+\tau=l \Delta t$. }
    \label{fig:twotime}
\end{figure}

In practice, this is done by applying the operator $X$ to the waveguide bin at index $k$, corresponding to $t = k \Delta t$, and the operator $Y$ to the bin at index $l$, corresponding to $t+\tau = l \Delta t$, on the state obtained after a single time evolution. We illustrate this in Fig.~\ref{fig:twotime}. As for single-time expectation values, the tensors to the left and right of the relevant bins are contracted iteratively (see Fig.~\ref{fig:exp_val}(c)). In addition, the bins between indices $k$ and $l$ must be contracted, which can also be performed iteratively. The computation of $\expval{X(t) Y(t+\tau)}$ is often computationally more demanding than the time evolution itself, due to the large number of contractions required to have all combinations of the two times $t$ and $t+\tau$.

\section{Time-delayed feedback dynamics of a two-level system \label{sec:tls_nmark}}
\begin{figure}[!ht]
    \centering
    \includegraphics[width=\linewidth]{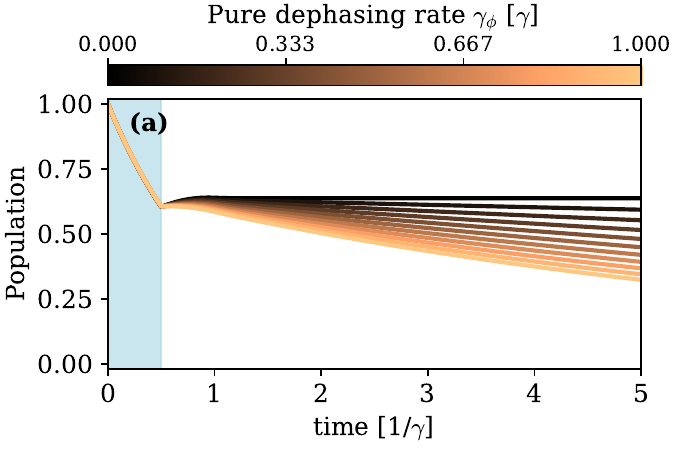}
    
    \vspace{0.5em}
    
    \includegraphics[width=\linewidth]{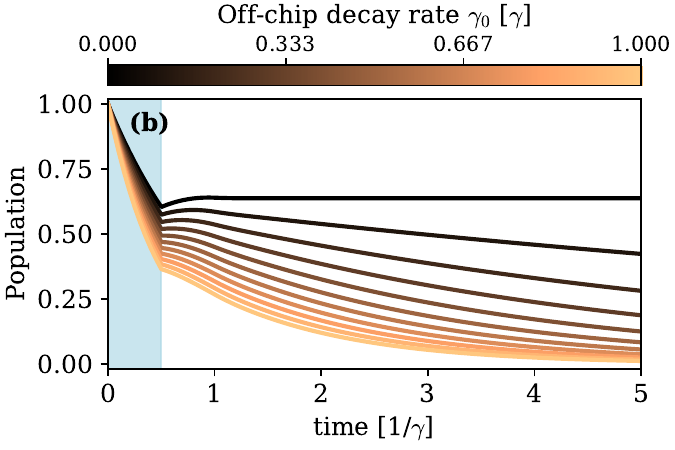}
    
    \caption{The population of a two-level system in a semi-infinite waveguide as a function of time with a mirror phase of $\phi=\pi$. In (a), we show curves for a varying pure dephasing rate, while in (b) we vary the off-chip decay rate $\gamma_{0}$, as denoted by the colorbars. The blue shaded area shows the time before the first delay $\tau$.}
    \label{fig:1tls_feedback}
\end{figure}

As a first demonstration of the method, we consider a single TLS coupled to a semi-infinite waveguide terminated by a mirror, again see Fig.~\ref{fig:sketch_tls_nmark} for an illustration. In this geometry, photons emitted into the waveguide are reflected by the mirror and return to interact with the emitter after a round-trip time $\tau$, leading to coherent {\it time-delayed feedback}, yielding non-Markovian dynamics. The mirror also imparts a phase $\phi$ to the reflected field, which depends on the emitter–mirror distance.

In the absence of decoherence, this system is known to exhibit population trapping when the feedback phase satisfies $\phi=\pi$. In this case, destructive interference between the emitted field and the returning photon suppresses spontaneous emission, resulting in a long-lived excited-state population \cite{Carmele2013,ArranzRegidor2021CavitylikeRegime,Bundgaard-Nielsen2024WaveguideQED.jl:Electrodynamics}.

Here, we investigate how this behavior is modified in the presence of decoherence. We consider both pure dephasing and decay outside the waveguide mode (off-chip decay). The pure dephasing is included through the Lindblad dissipation term $\gamma_\phi \mathcal{D}_{\sigma^\dagger \sigma}[\rho]$ in the system Liouvillian in Eq.~\eqref{eq:rho_eom}, where $\gamma_\phi$ denotes the dephasing rate, and $\mathcal{D}_A(\rho)=A \rho A^{\dag}-\frac{1}{2}\left(A^{\dagger} A \rho+\rho A^{\dagger} A\right)$ is the Lindblad superoperator. Similarly, off-chip decay can be included as $\gamma_{0} \mathcal{D}_{\sigma}[\rho]$. We sweep both decoherences from $\gamma_{\phi/0} = 0$ to $\gamma_{\phi/0} = \gamma$, where $\gamma$ is the total decay rate into the waveguide [see Eq.~\eqref{eq:tls_nmar_hamiltonian}].

The resulting population dynamics of the emitter are shown in Fig.~\ref{fig:1tls_feedback}, where we consider a 
feedback delay of $\tau = 0.5/\gamma$, see Appendix \ref{app:numerical_details} for numerical details. In the absence of decoherence, the excited-state population remains trapped due to the destructive interference created by the delayed feedback. In Fig.~\ref{fig:1tls_feedback}(a), as the pure dephasing rate increases, this interference is gradually disrupted, allowing the photon to pass the emitter and thus leading to an eventual slow decay of the system.

Notably, the effect of feedback remains clearly visible even for strong dephasing. For example, when $\gamma_\phi = \gamma$, the population after ten round-trip times is still approximately half of the value obtained in the lossless case. This demonstrates that the feedback-induced trapping remains partially robust against decoherence even when the pure dephasing rate is on the same order of magnitude as the emitter-decay rate.

As for off-chip decay, shown in Fig.~\ref{fig:1tls_feedback}(b), we observe a significantly faster decay that sets in already before the first delay time $\tau$. This is expected, as photons are now irreversibly lost to modes outside the waveguide. In contrast, pure dephasing does not directly remove photons, but instead perturbs the phase coherence required for the perfect destructive interference, thereby only indirectly leading to losses at later times. 

Both of the results in Fig.~\ref{fig:1tls_feedback}(a) and (b) reproduce results obtained with a discrete waveguide model using quantum trajectories \cite{ArranzRegidor2021ModelingModel}. With the density matrix MPS formulation presented here, however, we are not limited in the number of quanta, which can be increased beyond what is accessible in this discrete waveguide model.

 Here, we demonstrated that our approach can tackle any decoherence mechanism that can be included via Lindblad loss terms. While the effect of off-chip decay is important, its effects are more well studied, and as discussed above, can in principle be included in most pure state MPS methods by the addition of another waveguide mode. For these reasons, we study only the effects of pure dephasing in the following examples, although the inclusion of off-chip decay is easily possible.

\section{Time-delayed superradiance \label{sec:2tls}}

\begin{figure}[!ht]
    \centering
    \includegraphics[width=\linewidth]{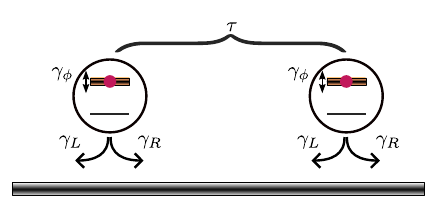}
    \caption{A sketch of two emitters coupled to the same waveguide with a separation of $\tau$. Both emitters experience the same total decay rate $\gamma$ and pure dephasing rate $\gamma_\phi$.
     }
    \label{fig:2ls_sketch}
\end{figure}

\begin{figure}
    \centering
    \includegraphics[width=\linewidth]{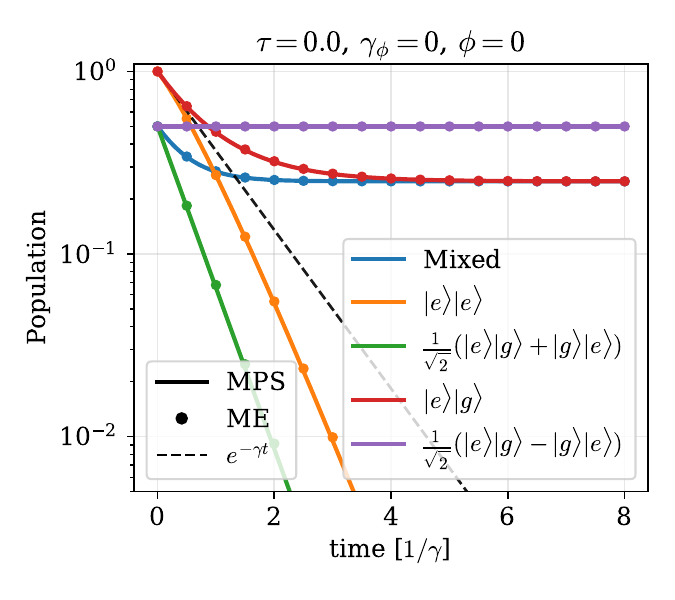}
    \caption{The time evolution of the first of the two TLSs coupled via the waveguide. The time delay $\tau$ between the emitters is here 0, meaning the dynamics is markovian. We show results obtained using the MPS method and the ME in Eq.~\eqref{eq:me} for different initial conditions, showing super and sub-radiant decays. The mixed state considered is given in Eq.~\eqref{eq:mixedstate} For reference, we also plot a dashed line for the exponential decay of $\exp(-\gamma t)$.
    }
    \label{fig:2lsmarkov}
\end{figure}

\begin{figure*}[!ht]
    \centering
    \includegraphics[width=\linewidth]{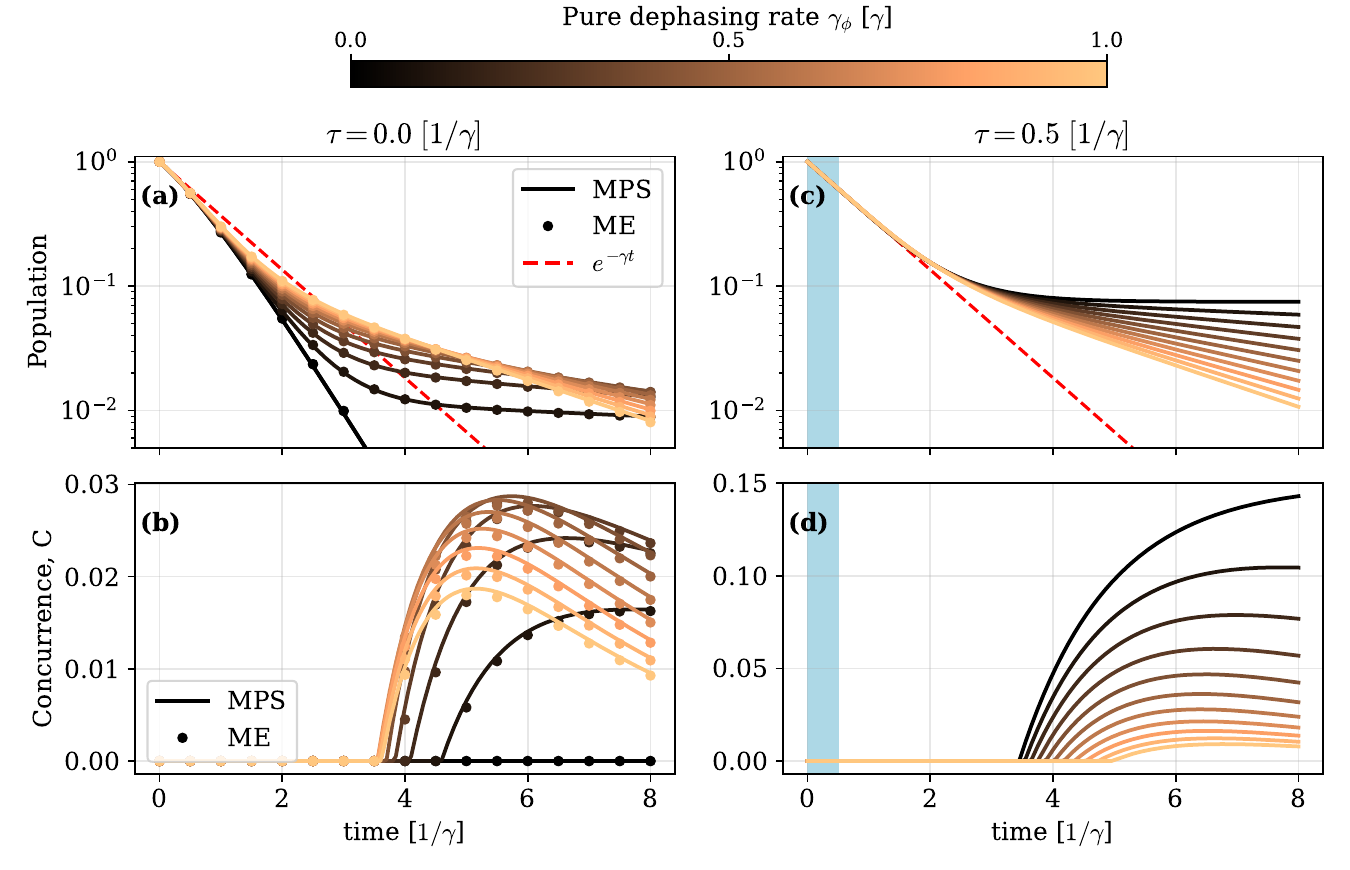}
    \caption{(a) and (b): The population of the left emitter separated from the right emitter by delays $\tau=0$ and $\tau=0.5 \ [1/\gamma]$, respectively. Both emitters are initially excited. For $\tau=0$, we also show the population as predicted by the ME in Eq.\eqref{eq:me}. (c) and (d), same as (a) and (b), but instead, the entanglement, defined as the concurrence $C$ in Eq.~\eqref{eq:concurrence} as a function of time. In all plots, the pure dephasing rate of both emitters is swept as denoted by the colorbar, and the phase difference between the emitters is $\phi=0$. The blue shaded area shows the time before the first delay $\tau$.\label{fig:two_tls_main}  }
    
\end{figure*}

We now consider two TLSs coupled to the same waveguide and separated by a 
finite propagation delay $\tau$. Photons emitted by one emitter thus propagate through the waveguide and interact with the second emitter after the delay time $\tau$. See Fig.~\ref{fig:2ls_sketch} for a sketch of such a system. This waveguide-mediated interaction leads to collective emission effects that depend on both the emitter separation and the coherence of the emitters. By considering two waveguide modes and two emitters and following the derivation in Sec.~\ref{sec:timebin}, one can arrive at the Hamiltonian \cite{ArranzRegidor2021ModelingModel,ArranzRegidor2025TheoryExcitations}:
\begin{equation}
\begin{aligned}
    H_{k} &= \frac{1}{\sqrt{ \Delta t}}  \bigg [ ( \sqrt{\gamma_L} \mathrm{e}^{ i \phi} b_{L,k-\tilde \tau}  +  \sqrt{\gamma_R}  b_{R,k} ) \sigma_1^\dagger + {\rm H.c}    \\ 
    &+( \sqrt{\gamma_L} b_{L,k} +  \sqrt{\gamma_R} \mathrm{e}^{ i \phi} b_{R,k-\tilde \tau}  ) \sigma_2^\dagger + {\rm H.c} \bigg] ,
\end{aligned}
\end{equation}
where the subscript $b_{L/R,k}$, denotes the left/right going mode in the $k$'th bin and $\sigma_i$ denotes emitter $i$. The rates $\gamma_{L/R}$ denote the decay rate of the emitters into the left/right going mode. In the following, we consider symmetrical coupling $\gamma_L = \gamma_R = \gamma/2$, where $\gamma$ is then the total decay rate of the emitters. The phase $\phi$ here denotes the phase the waveguide field accrues during the propagation to the other emitter. 

\subsection{Delay and Entanglement between emitters}

If we have no delay between the emitters, $\tau = 0$, the dynamics are Markovian and can be accurately modeled using a Markovian master equation (ME). By tracing out the waveguide field, one arrives at the dynamics of the two emitters, which are governed by \cite{Pichler2016PhotonicFeedback,PhysRevA.66.063810,Caneva2015QuantumFormalism,PhysRevA.91.051803, Arranz2023Probingpumping}:
\begin{equation}
\begin{aligned}
    \frac{d \rho_S}{dt} &= -i[H_{\rm eff},\rho_S] + \frac{\gamma}{2} \mathcal{D}_{C_-}[\rho_S] + \frac{\gamma}{2} \mathcal{D}_{C_+}[\rho_S] \\ 
    & + \gamma_\phi \mathcal{D}_{\sigma_1^\dagger \sigma_1}[\rho_S]+ \gamma_\phi \mathcal{D}_{\sigma_2^\dagger \sigma_2}[\rho_S] \label{eq:me}
\end{aligned}
\end{equation}
with $C_{\pm} =\sigma_1 + \mathrm{e^{\pm i \phi}}\sigma_2 $ and $H_{\rm eff} = \frac{\gamma}{2}\mathrm{Im}\{\mathrm{e}^{i \phi}\}(\sigma_1^\dagger \sigma_2 + \sigma_2^\dagger \sigma_1)$. With this, we can test the MPS implementation in the appropriate limit (e.g., Markovian regime), and also showcase the limitations of the ME approach.

In Fig.~\ref{fig:2lsmarkov}, we show the population dynamics of emitter one for different initial states of the two-emitter system without any delays, $\tau=0$ (again for numerical details of the MPS see appendix \ref{app:numerical_details}). We first consider the case without pure dephasing, $\gamma_\phi = 0$, and compare the MPS results with those obtained from the master equation (ME).
Note that the ME makes both a Markov approximation and a Born approximation, so entanglement effects between the emitters and waveguide modes are neglected in such an approach.
Depending on the initial state, the system exhibits different decay behaviors, but in all cases the MPS and ME agree exactly (with no delay between emitters). 

For the initial superradiant state $\rho_S = \ket{\psi}\bra{\psi}$, with $|\psi\rangle = \frac{1}{\sqrt{2}}\left(|eg\rangle + |ge\rangle\right)$, we observe enhanced collective emission, where the decay occurs at twice the individual emitter rate, $2\gamma$. A similar enhancement is observed when both emitters are initially excited, $\ket{\psi} = \ket{ee}$. In this case, the system exhibits accelerated emission with an effective decay rate of $\sqrt{2}\gamma$ \cite{ArranzRegidor2025TheoryExcitations}.

The density-matrix MPS formulation also allows us to consider mixed initial states. We therefore study the statistical mixture example,
\begin{equation}
    \rho_S = \frac{1}{2}\left( \ket{g}\bra{g} \otimes \ket{e}\bra{e} + \ket{e} \bra{e} \otimes  \ket{g}\bra{g}\right), \label{eq:mixedstate}
\end{equation}
which has the same excitation probability as the superradiant state but lacks coherence between the emitters. As shown in Fig.~\ref{fig:2lsmarkov}, this absence of coherence suppresses superradiant behavior. Instead, the system evolves into a long-lived state with reduced decay.

This can be understood by noting that the state $\ket{eg}$ can be decomposed as:
\begin{equation}
    \ket{eg} = \frac{1}{\sqrt{2}}\left(\frac{1}{\sqrt{2}}\left(|eg\rangle + |ge\rangle\right) + \frac{1}{\sqrt{2}}\left(|eg\rangle - |ge\rangle\right) \right),
\end{equation}
corresponding to equal contributions from the superradiant and subradiant states. While the superradiant component decays rapidly, the subradiant component remains dark and therefore leads to a persistent population at long times. This behavior is also illustrated in Fig.~\ref{fig:2lsmarkov}, where the purely subradiant state $\frac{1}{\sqrt{2}}\left(|eg\rangle - |ge\rangle\right)$ shows no decay.

From all these observations, we see that the waveguide-mediated interaction between the two emitters is very rich even in the Markovian limit. Still, the MPS method allows us to investigate the impact of a finite time delay between the emitters, while also including pure dephasing into the model.

\subsection{Superradiance with a delay of $\tau=0.5/\gamma$}

In Fig.~\ref{fig:two_tls_main}, we consider two initially excited emitters, $\ket{ee}$, for delays $\tau = 0$ and $\tau = 0.5/\gamma$, see appendix \ref{app:numerical_details} for numerical details. In each case, the pure dephasing rate is varied, as indicated by the colorbar. The first row shows the population of the left emitter.

For zero delay, $\tau=0$, pure dephasing suppresses the coherence between the emitters and thus diminishes the superradiant effect. As the dephasing increases, the dynamics transition to a much slower decay. This behavior can be understood from the formation of a subradiant state. The pure dephasing breaks the symmetry of the emitters and allows a coupling between the superradiant and subradiant states. The subradiant state eventually also couples with the superradiant state and therefore decays slowly over time.

For a finite delay of $\tau = 0.5/\gamma$, the effect of the delay is already visible in the absence of dephasing. Initially, the emitters decay at the individual rate $\gamma$ until the delay time is reached, after which the dynamics deviate from this behavior. In contrast to the Markovian case, we do not observe a superradiant decay, but instead the emitters evolve towards a subradiant or trapped state. This behavior is similar to the trapping seen in Fig.~\ref{fig:1tls_feedback}, where destructive interference prevents the excitations from escaping. 

Without dephasing, this corresponds to the formation of a dark state between the spatially separated emitters. Introducing pure dephasing leads to a gradual decay of this long-lived correlation.

Another way to characterize correlations {\it between the emitters} is through entanglement measures. For a bipartite system composed of subsystems $A$ and $B$, the entanglement entropy is commonly defined as \cite{Nielsen_Chuang_2010}
\begin{equation}
    S = -\mathrm{tr}_A(\rho_A \log \rho_A) = -\mathrm{tr}_B(\rho_B \log \rho_B), \label{eq:entropy_ME}
\end{equation}
where $\rho_{A/B} = \mathrm{tr}_{B/A}(\rho)$ is the reduced density matrix of subsystem $A/B$. This quantity provides a meaningful measure of entanglement only when the total state \mbox{$\rho = \ket{\psi}\bra{\psi}$} is pure. In that case, the entropy $S$ quantifies the entanglement between the two subsystems.

For mixed states, however, the situation is more subtle. In particular, Eq.~\eqref{eq:entropy_ME} can yield a nonzero value even for purely statistical mixtures, and therefore does not reliably quantify entanglement. This issue has motivated the introduction of alternative entanglement measures for mixed states, such as the logarithmic negativity \cite{Vidal2002ComputableEntanglement, Plenio2005LogarithmicConvex} and the concurrence \cite{Wootters1998EntanglementQubits}.

The logarithmic negativity is relatively straightforward to compute and provides a sufficient condition for entanglement. The Concurrence, on the other hand, is closely related to the entanglement of formation and quantifies the minimal entanglement required to prepare a given state. While the general definition of Concurrence involves an optimization over all possible decompositions of the density matrix, for the special case of two qubits, an explicit formula was derived by Wootters \cite{Wootters1998EntanglementQubits}. In this case, the Concurrence is given by
\begin{equation}
    C(\rho)=\max \left\{0, \lambda_1-\lambda_2-\lambda_3-\lambda_4\right\}, \label{eq:concurrence}
\end{equation}
where $\lambda_i$ are the eigenvalues, in descending order, of the matrix $R = \sqrt{\sqrt{\rho} \tilde{\rho} \sqrt{\rho}}$, with $\tilde \rho = (\sigma_y \otimes \sigma_y) \rho^* (\sigma_y \otimes \sigma_y)$. 

In the lower panels of Fig.~\ref{fig:two_tls_main}, we show the Concurrence as a function of time for the two delays $\tau=0$ and $\tau = 0.5/\gamma$. In the Markovian case, no entanglement is generated in the absence of pure dephasing. Interestingly, the introduction of dephasing, which perturbs the superradiant state, leads to a small but finite amount of entanglement.

In the delayed case, we observe a significant buildup of entanglement between the two emitters. Notably, this entanglement only becomes non-zero after several feedback loops, appearing at times $t \approx 5\tau$–$6\tau$. In the absence of dephasing, the entanglement appears long-lived, which can be understood from the formation of a subradiant (dark) state, as discussed above and in Ref.~\cite{Alvarez-Giron2024Delayatoms}.

With the inclusion of pure dephasing, which to our knowledge has not been considered in this context before, the Concurrence is reduced and decays in the long-time limit. Nevertheless, even for $\gamma_\phi = 0.5\gamma$, a finite Concurrence persists at long times. Compared with the equivalent Markovian case, this long-time value is larger by approximately a factor of two, reflecting the formation of a delayed dark component. Thus, the finite delay modifies the mechanism by which emitter--emitter entanglement is generated and sustained, although pure dephasing still degrades the correlations over time.

\begin{figure}
    \centering
    \includegraphics[width=\linewidth]{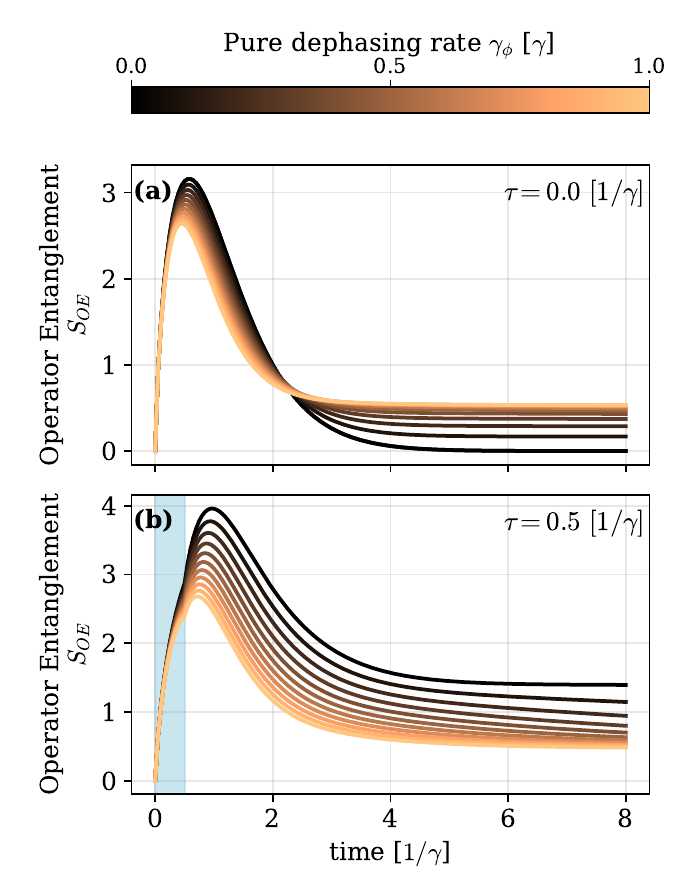}
    \caption{The operator entanglement $S_{OE}$ between the two emitters and the outgoing waveguide fields as a function of time in (a) for $\tau=0$ and in (b) for $\tau=0.5/\gamma$. The blue shaded area shows the time before the first delay $\tau$.
    }
    \label{fig:operatorentanglement}
\end{figure}

With the MPS method, it is also possible to characterize correlations between the emitters and the emitted waveguide field, which are not accessible within the ME description, where the field degrees of freedom have been eliminated. For a pure joint state of system and waveguide, the entanglement entropy can be obtained from the Schmidt decomposition
\begin{equation}
    \ket{\psi} = \sum_i \alpha_i \ket{u_i} \otimes \ket{v_i},
\end{equation}
with Schmidt coefficients $\alpha_i$, leading to the expression~\cite{Pichler2016PhotonicFeedback,ArranzRegidor2021ModelingModel}:
\begin{equation}
    S = - \sum_i \alpha_i \log \alpha_i.
\end{equation}

For mixed states, however, as discussed above, this quantity is no longer a faithful measure of entanglement. Applying a singular value decomposition to the density matrix, as in Eq.~\eqref{eq:svd_dm}, instead yields the {\it operator entanglement}~\cite{Wellnitz2022RiseDephasing,Daraban2025Non-unitarityDynamics},
\begin{equation}
    S_{OE} = - \sum_i \lambda_i \log \lambda_i,
\end{equation}
where $\lambda_i$ are the singular values of the density matrix. For pure states, the operator entanglement is directly related to the entanglement entropy by $S_{OE} = 2S$~\cite{Dubail2017Entanglement1+1d}.  For mixed states, however, $S_{OE}$ is instead a measure of the complexity of the density matrix and, correspondingly, of the MPS representation, rather than a direct measure of physical entanglement. A finite value of $S_{OE}$ means that the density operator does not factorize as a simple product across the chosen bipartition. This non-factorization can arise from genuine quantum entanglement, but it can also arise from statistical mixtures in the density matrix. Thus, while a product density matrix has zero operator entanglement, a mixed but separable state may still have a nonzero $S_{OE}$. In this sense, $S_{OE}$ should be interpreted as a complexity measure, not as an entanglement monotone for mixed states.

In Fig.~\ref{fig:operatorentanglement}, we show the operator entanglement $S_{OE}$ obtained from the MPS method, where the system is bipartitioned into the two emitters and the outgoing waveguide field, for varying dephasing rates and for delays $\tau=0$ and $\tau=0.5/\gamma$. 

For $\tau=0$ and $\gamma_\phi = 0$, the operator entanglement decays to zero in the long-time limit, consistent with the expectation that the total state becomes separable once the emitters have fully decayed. For $\gamma_\phi \neq 0$, however, we observe that $S_{OE}$ approaches a finite, nonzero value even at long times. This indicates that the joint system–waveguide state does not factorize into a simple product state, but instead retains a nontrivial structure. Importantly, this should not be interpreted as persistent physical entanglement, but rather as a signature of residual correlation or the presence of a statistically mixed density matrix, which requires a finite bond dimension to represent.

Introducing a finite delay leads to a more complex evolution of the operator entanglement, as well as generally larger values, reflecting the increased structure induced by time-delayed feedback. Interestingly, while for $\tau=0$ the long-time value of $S_{OE}$ increases with the dephasing rate, for $\tau=0.5/\gamma$ we observe the opposite behavior. In this case, increasing $\gamma_\phi$ reduces $S_{OE}$, indicating that pure dephasing suppresses correlations between the emitters and the emitted field in the presence of feedback.

We note that while we here only studied the operator entanglement, other entanglement measures between the emitter and waveguide field are available. For example, the log-negativity \cite{Vidal2002ComputableEntanglement,Plenio2005LogarithmicConvex}, which would not be computable in a ME approach, even in the Markovian limit, due to the elimination of the waveguide field.

In the next section, we investigate further the impact of pure dephasing on the out-coupled field.

\section{Emission spectra and emitter excitation with Fock state pulses \label{sec:fockpulses}}

\begin{figure}[!ht]
    \centering
    \includegraphics[width=\linewidth]{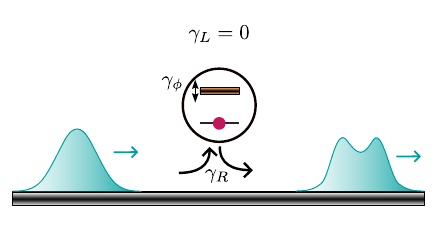}
    \caption{Sketch of a one- or two-photon pulse scattering off a chirally coupled emitter (TLS), meaning the scattered pulse only goes to the right.}
    \label{fig:pulse_sketch}
\end{figure}

As a final example, we consider a chirally coupled TLS driven by a quantized input pulse under the effect of pure dephasing as illustrated in Fig.~\ref{fig:pulse_sketch}. This problem has already been studied without decoherence in Ref.~\cite{ArranzRegidor2025TheorySpectra}, and we use it here as an application of our pure dephasing framework. In that work, it was shown that a single-photon pulse scattering off a chirally coupled two-level system is transmitted without any changes in its spectrum.
Specifically, this is the stationary or long-time spectrum, 
as there is also a dynamical spectrum with pulsed excitation
\cite{ArranzRegidor2025TheorySpectra}.
This occurs despite nontrivial dynamics of the emitter during the interaction, reflecting the effectively linear response of the system in the single-excitation manifold (coinciding with a bosonic regime). However, introducing two excitations breaks this linear regime, and thus the scattered two-photon pulse is distorted by the nonlinear interaction with the emitter. 

In the following, we will see that the pure dephasing can also serve as a way to break the linear (bosonic) regime and cause an effective distortion of the scattered pulse, regardless of whether it contains one or two photons. We also believe such symmetry breaking is experimentally accessible with current waveguide QED platforms, especially quantum dots, where single-photon Fock state pulses are readily available.

\subsection{One- and two-photon pulse states
with MPS}

To begin with, we discuss how to represent Fock states in the MPS picture using kets, before generalizing to density matrices. In terms of the discretized operators, we write a one-photon Fock state as
\begin{equation}
    \ket{\psi} = \sum_k \sqrt{\Delta t} f_k b_k^\dagger \ket{0} ,\label{eq:singlephotonstate}
\end{equation}
where $f_k = f(t_k)$ is the pulse envelope function, in discretized time. As discussed in Sec.~\ref{sec:timebin}, the state in Eq.~\eqref{eq:singlephotonstate} is too large to numerically represent, which also means that we cannot sequentially decompose it using SVDs. The state is, however, simple enough that an analytical MPS decomposition of Eq.~\eqref{eq:singlephotonstate} exists \cite{Barkemeyer2021StronglyState,ArranzRegidor2025TheorySpectra}. We write the MPS decomposition as 
\begin{equation}
    \ket{\psi}_{ij\cdots k} = A^1_{i\alpha} A^2_{\alpha j \beta} \cdots A^N_{\gamma k}
\end{equation}
where $\ket{\psi}_{ij\cdots k}$ denotes the $i$'th element of the first site, the $j$'th of the second, and so on, and where Einstein summation over repeated indices is assumed. The tensors here have bond dimension $\chi=2$. 

For the first edge, we have
\begin{equation}
    A^1_{:,0} = (1,0),
    \qquad
    A^1_{:,1} = (0,f_1),
\end{equation}
where the colon denotes the physical index of the local Hilbert space.

The subsequent tensors have the blocks given by
\begin{equation}
    A^k_{0,:,0} = (1,0),
    \qquad
    A^k_{0,:,1} = (0,f_k),
\end{equation}
\begin{equation}
    A^k_{1,:,1} = (1,0),
    \qquad
    A^k_{1,:,0} = (0,0),
\end{equation}
Finally, the last edge is given by
\begin{equation}
    A^N_{0,:} = (1,0),
    \qquad
    A^N_{1,:} = (0,f_N).
\end{equation}

The interpretation is as follows and similar although formulated differently, from the one presented in Ref.~\cite{Crosswhite2008}: the first bond index ($0$) denotes that no excitation has been placed so far in the MPS chain. Thus, $A^k_{0,:,0} = (1,0)$ represents the vacuum state. A transition from bond index $0$ to $1$ via $A^k_{0,:,1} = (0,f_k)$ places the excitation in the $k$'th bin. The block $A^k_{1,:,1} = (1,0)$ then represents the vacuum state after the excitation has already been placed in the chain. The transition from bond index $1$ back to $0$ is not allowed and therefore vanishes.

\begin{figure}[!ht]
    \centering
    \includegraphics[width=\linewidth]{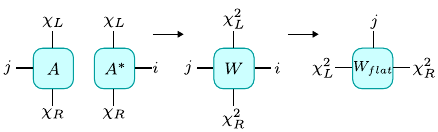}
    \caption{Sketch of the local tensor product of a ket-tensor $A$ to the density tensor $W$ as displayed in Eqs.\eqref{eq:A_in}-\eqref{eq:W_flat}}
    \label{fig:ket_to_dm}
\end{figure}

\begin{figure*}[!ht]
    \centering
    \includegraphics[width=\linewidth]{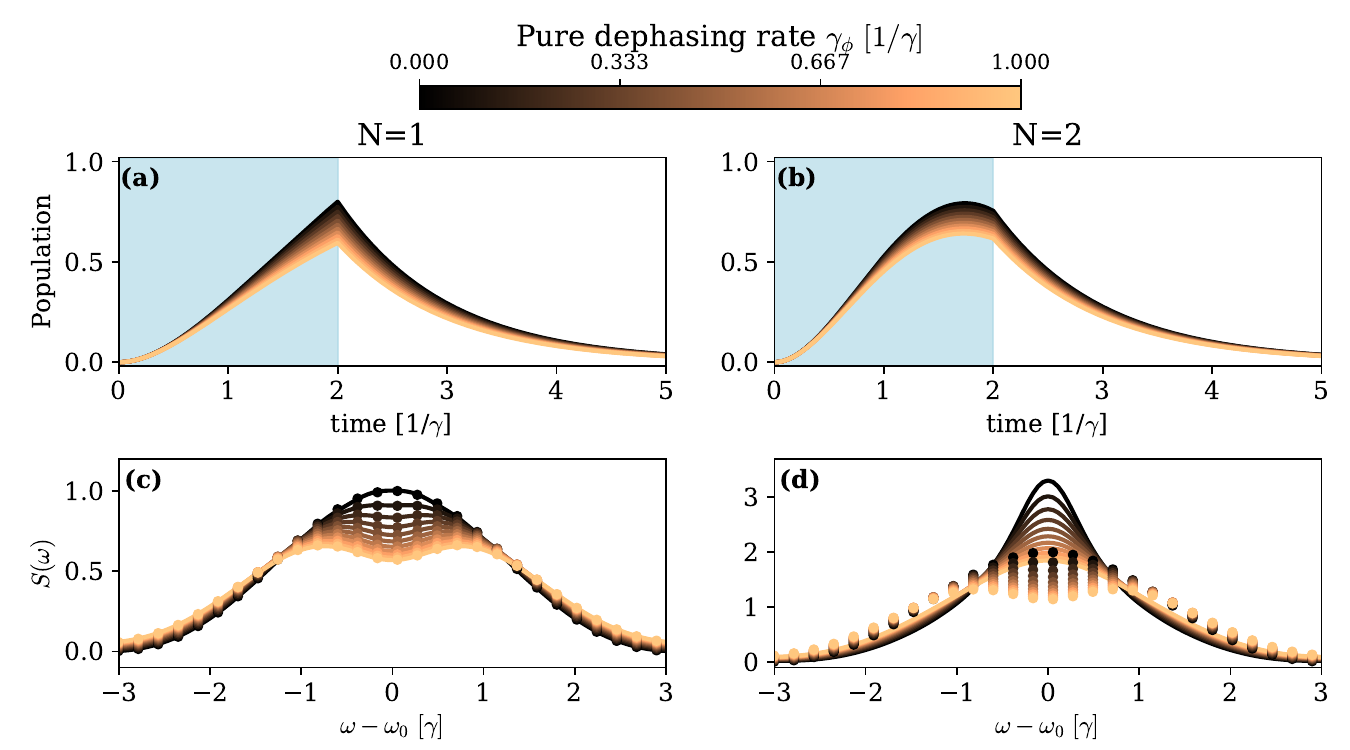}
    \caption{(a) and (b): The population as a function of time for a top-hat input pulse of $N=1$ and $N=2$ photons, respectively. Pulse duration is $T_P = 2 /\gamma$ as also indicated by the blue shaded area. The pure dephasing rate is varied as indicated by the colorbar. (c) and (d): same as above, but instead the emission spectrum $S(\omega)$ in Eq.\eqref{eq:spec}. The analytically predicted spectrum for a single photon pulse, including pure dephasing in Eq.~\eqref{eq:s_total} is also shown in full circles.
    }
    \label{fig:spectrum}
\end{figure*}

For a two-photon Fock state, which we write as
\begin{equation}
    \ket{\psi} = \frac{1}{\sqrt{2}} \sum_{k,k'} \Delta t\, f_k f_{k'}\, b_k^\dagger b_{k'}^\dagger \ket{0},
\end{equation}
we now need a bond dimension of $\chi=3$ and a local Hilbert space dimension of $3$ (vacuum, one-, or two-excitation states). The decomposition follows the same structure as in the one-photon case, but the bond indices now track whether zero, one, or two excitations have been placed in the chain. For the bulk tensors, the nonzero blocks are \cite{Barkemeyer2021StronglyState,ArranzRegidor2025TheorySpectra}
\begin{align}
    &A^k_{0,:,0} = (1,0,0), \\
    &A^k_{0,:,1} = (0,f_k,0), \\
    &A^k_{0,:,2} = (0,0,f_k^2/\sqrt{2}),
\end{align}
\begin{equation}
    A^k_{1,:,1} = (1,0,0), \qquad
    A^k_{1,:,2} = (0,f_k,0),
\end{equation}
\begin{equation}
    A^k_{2,:,2} = (1,0,0),
\end{equation}
with all other blocks vanishing. Note that other decompositions exist due to gauge freedom. The interpretation is a direct extension of the one-photon case. The bond indices $0,1,2$ denote that zero, one, or two excitations have been placed in earlier bins. Transitions $0\rightarrow1$ and $1\rightarrow2$ correspond to placing the first and second excitation, respectively, while the diagonal blocks connect vacuum states. The block $A^k_{0,:,2}$ corresponds to placing two excitations in the same time bin. As before, transitions that would decrease the bond index are not allowed. Similar logic can be followed for $N$-photon Fock states \cite{regidor2026quantumdynamicsfewphotonpulsed}. 

The extension from a pure-state MPS to a density-matrix representation is straightforward for an initial state of the form $\rho = \ket{\psi}\bra{\psi}$. If a local MPS tensor is written as
\begin{equation}
    A \in \mathbb{C}^{\chi_L \times d \times \chi_R} \, , \label{eq:A_in}
\end{equation}
then the corresponding local density-matrix tensor is constructed site-wise as
\begin{equation}
    W_{(a,a'),\,i,\,j,\,(b,b')} = A_{a i b} A^{*}_{a' j b'} ,
\end{equation}
which is simply the local tensor product $A \otimes A^*$. Here $i$ and $j$ denote the ket and bra physical indices, respectively, while $(a,a')$ and $(b,b')$ are the corresponding fused left and right bond indices. The resulting tensor, therefore, has dimensions
\begin{equation}
    W \in \mathbb{C}^{\chi_L^2 \times d \times d \times \chi_R^2}.
\end{equation}

Since we work with vectorized density matrices, the two physical indices are then flattened into a single index, yielding
\begin{equation}
    W_{\mathrm{flat}} \in \mathbb{C}^{\chi_L^2 \times d^2 \times \chi_R^2}. \label{eq:W_flat}
\end{equation}
Consequently, an initial pure-state MPS can be converted directly into the corresponding density-matrix MPS used in the density matrix code. This process is also illustrated in Fig.~\ref{fig:ket_to_dm} and in Appendix \ref{app:dm_mps_pulses}, we explicitly calculate and interpret the density matrix MPS decomposition tensors.

\subsection{Scattering of few-photon Fock pulses with a chiral TLS \label{sec:chiral_tls}}

With the MPS representation of one- and two-photon Fock states, we can now study the scattering of a chirally coupled TLS. In the following, we consider either a one-photon or a two-photon Fock state with a top-hat envelope function (this is not a model restriction, and any arbitrary shape can be used), defined as
\begin{equation}
    f(t) = \frac{1}{\sqrt{T_p}}, \qquad 0\leq t < T_p,
\end{equation}
and zero otherwise. We consider a pulse duration of $T_p = 2/\gamma$.

To investigate the impact of pure dephasing on the scattered pulse, we consider the time-integrated 
or stationary spectrum:
\begin{equation}
    S(\omega) = \int_0^\infty dt \int_0^\infty d\tau \, \mathrm{e}^{i\omega \tau} \expval{b_R^\dagger(t) b_R(t+\tau)}, \label{eq:spec}
\end{equation}
where $b_R$ is the annihilation operator of the right-propagating waveguide mode. For the initial top-hat pulse, we have $\expval{{b_{R}^{\dagger\rm in} }(t) b_R^{\rm in}(t+\tau)} = N f(t)f(t+\tau)$ where $N=1$ for a single-photon pulse and $N=2$ for a two-photon pulse; using this, the input spectrum can be calculated as
\begin{equation}
    S^{\rm in}(\omega) = N \frac{\sin^2(\omega T_p/2)}{(\omega T_p/2)^2}, \label{eq:spec_initial}
\end{equation}
where the superscript "${\rm in}$" denotes the input spectrum.

In the absence of detuning and pure dephasing and for a single-photon input, the spectrum remains unchanged after scattering. This result was studied in detail in Ref.~\cite{ArranzRegidor2025TheorySpectra}, where it was shown analytically that the emitter dynamics conspire to exactly cancel out when the spectrum is integrated over time, leaving the pulse only with a phase-change that does not show up in the spectrum. In Appendix \ref{app:chiral}, we derive an extension of the results to include pure dephasing for the single photon scattering. The pure dephasing leads to incoherent emission by the emitter, and there is no longer a cancellation of the scattered and emitted field of the emitter, leading to a change in the output spectrum. The emission spectrum is given as:
\begin{equation}
    S(\omega) = S_{\rm coh}(\omega) + S_{\rm inc}(\omega), \label{eq:s_total}
\end{equation}
where 
\begin{equation}
    S_{\rm coh}(\omega) = |f(\omega)|^2 |t(\omega)|^2, \ \ \ t(\omega)=1-\frac{\gamma}{\Gamma + i (\omega-\delta)},
\end{equation}
and $\Gamma=(\gamma_\phi + \gamma)/2$, $\delta=\omega_0-\omega_p$, and $\omega_p$ is the frequency of the pulse. 

We consider $\delta=0$ in the following. The expression for the coherent emission is well-known in the literature, see for example Ref.~\cite{Ramos2018}, where they include pure dephasing, but only calculate the coherently scattered part. The coherent part of the emission thus dips around the resonance, since $|t(0)|^2 =  (1-\frac{2\gamma}{\gamma+\gamma_\phi})^2$. As for the incoherent part, it has the expression:
\begin{equation}
    S_{\rm inc}(\omega) = \frac{2\gamma\gamma_\phi\Gamma}
{(\omega-\delta)^2+\Gamma^2}
\int\frac{d\nu}{2\pi}
\frac{|f(\nu)|^2}
{(\nu-\delta)^2+\Gamma^2},
\end{equation}
which has a Lorentzian-like shape around the resonance and disappears for $\gamma_\phi \rightarrow 0$. Thus, in the total spectrum, for pure dephasing $\gamma_\phi \neq 0$, we would expect a Lorentzian-like shape with a dip in the middle.

Figure~\ref{fig:spectrum}, panels (a) and (b), show the excitation probability of the TLS as a function of time for one- and two-photon pulses, respectively. Numerical details are provided in Appendix \ref{app:numerical_details}. The dynamics differ significantly between the two cases, reflecting the nonlinear response of the TLS. Varying the pure dephasing rate, we observe that dephasing reduces the maximum achievable excitation probability.

Figure~\ref{fig:spectrum}, panels (c) and (d), show the emitted spectrum after scattering, again for one- and two-photon pulses, respectively. As a reference, we plot the analytical expression in Eq.~\eqref{eq:s_total} in circles. For the single-photon case without dephasing, the spectrum remains unchanged, as expected. However, as the dephasing rate increases, the spectrum is modified and adopts a non-trivial shape: it is not simply reduced in magnitude or broadened, but rather two peaks occur for large pure dephasing values. This is predicted both by the analytical expression and by the MPS simulations. To our knowledge, this effect has not been previously reported, as it requires the inclusion of pure dephasing, which has not been widely considered in theoretical treatments. As mentioned above, the physical intuition is that pure dephasing leads to both coherent and incoherent emission, and these two no longer interfere in such a way that the spectrum is unchanged.

For the two-photon case, the scattered spectrum is changed even in the absence of pure dephasing due to the nonlinear interaction with the TLS. Increasing the dephasing further leads to a broadening and flattening of the spectrum. The analytical expression in Eq.~\eqref{eq:s_total} fails here, as this is only a linear treatment with no two-photon components. This highlights the difference between scattering single- and two-photon pulses from an emitter.

If one instead includes off-chip decay (not shown), similar features in the emission spectrum appear, but these features are also accompanied by an overall reduced magnitude. These differences can thus serve as a way to probe experimentally whether pure dephasing or off-chip decay is the dominant decoherence process.

\section{Outlook and Conclusions \label{sec:conclusion}}
We have presented a density-matrix matrix product state approach for simulating waveguide QED systems with decoherence processes, which are captured through Lindbladian superoperators. By extending the time-bin MPS framework to density matrices, we are able to not only include decoherence processes such as pure dephasing directly in the dynamics, but also study mixed initial states.

Using this approach, we have studied three representative systems with significant waveguide QED effects, all of which are non-trivial, even with pure state dynamics. For a single emitter with time-delayed feedback, we showed that pure dephasing disrupts the interference responsible for population trapping, leading to a slow decay of the excited state. For two spatially separated emitters, we demonstrated how pure dephasing suppresses superradiant behavior and modifies the buildup of correlations between the emitters. We also showed that quantities such as operator entanglement provide additional insight into the structure of the joint emitter–field state, which is not accessible in a Markovian master-equation description. Finally, for the scattering of few-photon Fock state pulses from a chirally coupled emitter, we found that pure dephasing leads to qualitative changes in the emitted spectrum, including features not present in the lossless case. Off-chip decay leads to qualitatively similar features, although the overall magnitude of the spectrum is reduced.

Our method provides a powerful framework for studying non-Markovian waveguide QED systems in the presence of decoherence while maintaining access to both system and field observables. This is expected to be relevant for more realistic modeling of experiments, where such decoherence processes are unavoidable. Examples of experiments on superconducting qubits where pure dephasing is required to explain the data that already exists \cite{2603.28004,Hoi2015,Cheng2025,Odeh2025,Mirhosseini2019,Ferreira2024}.

Possible extensions and interesting directions include incorporating more complicated noise models into the dynamics, for example, including the impact of emitter-phonon coupling, and exploring their interplay with waveguide-mediated dynamics. Numerically exact methods exploiting MPS with phonons already exist \cite{Strathearn2018EfficientOperators,Cygorek2024ACE:Tensors,Cygorek2024SublinearSimulations,Link2024OpenContraction,Pollock2018OperationalProcesses,Jrgensen2019ExploitingIntegrals}, and with the extension to density matrices in the current work, a combination of the two methods should be possible. This was explored in Ref.~\cite{PhysRevResearch.3.023168}, but numerical complexity limited the systems available for study. With recent progress in the efficiency of tensor network algorithms, extensions to more complicated systems might be possible \cite{Cygorek2024SublinearSimulations,Link2024OpenContraction}. Including pure dephasing in the study of the so-called ``decoherence-free'' interaction of giant atoms is also an interesting direction \cite{PhysRevLett.120.140404}, where the role of pure dephasing has not been studied, yet is known to be essential for explaining experimental data on practically all waveguide QED systems.

The code necessary to reproduce the MPS results presented in this paper is available in the open source Python package QwaveMPS~\cite{arranzRegidor2026Qwavemps}.

\begin{acknowledgments}
This work was supported by the Danish National Research Foundation through NanoPhoton - Center for Nanophotonics, Grant No. DNRF147;
the Natural Sciences and Engineering
Research Council of Canada (NSERC) [Discovery Grant and Quantum Alliance],  the Canadian Foundation
for Innovation (CFI), and Queen’s University, Canada.

\end{acknowledgments}

\appendix
\section{Density-matrix MPS representation of few-photon pulse states}
\label{app:dm_mps_pulses}

In this appendix, we derive the density-matrix matrix product state (MPS) representation corresponding to the few-photon pulse states introduced in Sec.~\ref{sec:fockpulses}. The construction follows directly from the pure-state MPS by forming the outer product $\rho = \ket{\psi}\bra{\psi}$ sitewise.

Given a local MPS tensor,
\begin{equation}
    A^k_{\alpha,i,\beta},
\end{equation}
with bond indices $\alpha,\beta$ and physical index $i$, the corresponding density-matrix tensor is constructed as
\begin{equation}
    A^k_{(\alpha,\bar{\alpha}),\,(i,j),\,(\beta,\bar{\beta})}
    =
    A^k_{\alpha,i,\beta}\,
    \big(A^k_{\bar{\alpha},j,\bar{\beta}}\big)^*,
    \label{eq:dm_tensor_general}
\end{equation}
where $(\alpha,\bar{\alpha})$ and $(\beta,\bar{\beta})$ denote fused bond indices, and $(i,j)$ are the ket and bra physical indices. In the main text, the physical indices are flattened such that $(i,j)\rightarrow \tilde{i}$ with dimension $d^2$.

\subsection{One-photon state}
For the one-photon pulse, the bond dimension of the pure-state MPS is $\chi=2$, and the corresponding density-matrix MPS therefore has a bond space of dimension $\chi^2=4$, spanned by the pairs:
\begin{equation}
    (0,0),\; (0,1),\; (1,0),\; (1,1).
\end{equation}
Here, the first index refers to the ket chain and the second to the bra chain, indicating whether the excitation has been placed in each, respectively.

In this representation, each local tensor $A^k_{(a,\bar{a}),:,(b,\bar{b})}$ is a $2\times2$ matrix acting on the local Hilbert space spanned by $\{|0\rangle,|1\rangle\}$, where the row and column indices correspond to the ket and bra indices, respectively. The nonzero bulk tensors are then given by:
\begin{equation}
\begin{aligned}
&A^k_{(0,0),:,(0,0)}=
\begin{pmatrix}
1 & 0\\
0 & 0
\end{pmatrix},
\qquad
A^k_{(0,0),:,(0,1)}=
\begin{pmatrix}
0 & f_k^*\\
0 & 0
\end{pmatrix},
\\
&A^k_{(0,0),:,(1,0)}=
\begin{pmatrix}
0 & 0\\
f_k & 0
\end{pmatrix}, 
\qquad 
A^k_{(0,0),:,(1,1)}=
\begin{pmatrix}
0 & 0\\
0 & |f_k|^2
\end{pmatrix},
\end{aligned} 
\end{equation}

\begin{equation}
A^k_{(0,1),:,(0,1)}=
\begin{pmatrix}
1 & 0\\
0 & 0
\end{pmatrix},
\qquad
A^k_{(0,1),:,(1,1)}=
\begin{pmatrix}
0 & 0\\
f_k & 0
\end{pmatrix},
\end{equation}

\begin{equation}
A^k_{(1,0),:,(1,0)}=
\begin{pmatrix}
1 & 0\\
0 & 0
\end{pmatrix},
\qquad
A^k_{(1,0),:,(1,1)}=
\begin{pmatrix}
0 & f_k^*\\
0 & 0
\end{pmatrix},
\end{equation}

\begin{equation}
A^k_{(1,1),:,(1,1)}=
\begin{pmatrix}
1 & 0\\
0 & 0
\end{pmatrix}.
\end{equation}
with all other blocks vanishing.

The interpretation follows directly from the pure-state construction: the bond indices track whether the photon has been placed in the ket and bra chains. The nonzero transitions, therefore, correspond to inserting the excitation in the ket, in the bra, or in both. In contrast to the pure-state case, the excitation can now be distributed across different sites, such that the ket and bra indices are excited at different positions. This is, for example, reflected in transitions from $(0,0)$ to $(0,1)$ at one site, followed by $(0,1)$ to $(1,1)$ at a later site.\\

\section{Numerical details\label{app:numerical_details}}
In this appendix, we provide the numerical details and runtimes for the simulations presented in the paper. The run times presented were run on a desktop PC with an AMD Ryzen 5 3600 CPU, 3.6 GHz, using 2 Cores. Although the MPS algorithm itself does not benefit from multi-threading, the linear algebra operations involved do, and thus running on more cores will provide a speedup, especially when larger bond dimensions are involved. The run times are not statistical averages and thus should be taken as rough estimates of the computational times, showing the efficiency of the simulations. Through all simulations, we kept the relative cutoff of the singular values to $\epsilon_{\rm rel} = 10^{-7}$  

For the non-Markovian TLS simulations performed in Fig.~\ref{fig:1tls_feedback}, we used a maximum bond dimension of $\chi = 4$ with a time step of $\Delta t = 0.05 / \gamma$. The run time for these simulations was $\approx 0.5 \ \rm s$ per pure dephasing simulation run. The trace and Hermiticity errors were both below $10^{-14}$.

For the superradiance simulations in Fig.~\ref{fig:2lsmarkov} and Fig.~\ref{fig:two_tls_main},
which is a particularly difficult example, we found that we needed a maximum bond dimension of $\chi = 40$, when both emitters were initially excited, and the pure dephasing was substantial. For fewer excitations or smaller pure dephasing values, the following simulation times will be faster, due to reduced complexity. Again, we used a time step of $\Delta t = 0.05 / \gamma$, and the computational time was $\approx 24 \ \rm min$ per pure dephasing value. The trace-preservation errors were here $\epsilon_{\rm tr} < 5 \cdot 10^{-3}$ and the Hermiticity violations $\epsilon_\mathrm{Herm} < 5 \cdot 10 ^{-6}$. Even though the trace preservation error is not extremely small, increasing the maximum bond dimension to $\chi = 80$, changed the maximum relative error in the population dynamics by $0.006$, and we thus consider the simulation converged.

For the scattering simulations in Fig.~\ref{fig:spectrum}, we used $\Delta t = 0.02/\gamma$, we used a maximum bond dimension of $\chi = 20$, and observed trace-preservation violations of $\epsilon_{\rm tr} < 1.6 \cdot 10^{-6}$ and the Hermiticity violations $\epsilon_\mathrm{Herm} < 2.15 \cdot 10 ^{-14}$. For the two-photon pulse, which has the largest initial bond dimension (9), the computational times were $\approx 15 \ \rm s$ per pure dephasing value. Notably, the time for computing the two-time average of $\braket{b^\dagger_R(t)b_R(t+\tau)}$ was subsequently $\approx 5 \ \rm min$ per pure dephasing value, substantially longer than the time evolution itself.

However, given the complexity and power of having an exact simulation method, with multiple quanta, non-Markovian feedback, and dissipation, these simulation times are still very reasonable and can be performed on a single computer workstation. With more cores on a high-performance cluster, even faster computation times are expected.

\section{Derivation of Single-Photon Stationary Spectrum Including Pure Dephasing \label{app:chiral}}
In Sec.~\ref{sec:chiral_tls}, we showed the stationary emission spectrum from a single photon pulse scattering off a chiral emitter. This configuration has already been studied in Ref.~\cite{ArranzRegidor2025TheorySpectra}, but without the inclusion of pure dephasing. In this appendix, we extend the analytical results to include pure dephasing and find that the spectrum can be separated into a coherent and an incoherent part.

Using the input-output relation, the output field can be written as (in the following we drop the subscript $R$ since we consider the chiral case and only one field exists) \cite{Gardiner1985InputEquation}:
\begin{equation}
    b_{\rm out}(t) = b_{\rm in}(t) - \sqrt{\gamma} \sigma(t).
\end{equation}

The output spectrum can therefore be written as:
\begin{equation}
\begin{aligned}
    S(\omega) &= \int_0^\infty dt \int_0^\infty d\tau \, \mathrm{e}^{i\omega \tau} \expval{b_{\rm out}^\dagger(t) b_{\rm out}(t+\tau)} \\ 
    &= \expval{b_{\rm out}^\dagger(\omega) b_{\rm out}(\omega)} \\
    &= \expval{{b_{\rm in}}^\dagger(\omega)b_{\rm in}(\omega)} - \sqrt{\gamma} \expval{b_{\rm in}^\dagger (\omega) \sigma(\omega)} \\ 
    &- \sqrt{\gamma} \expval{\sigma^\dagger(\omega) b_{\rm in}(\omega) } + \gamma \expval{\sigma^\dagger(\omega)\sigma(\omega)}. \label{eq:stationary_spectrum_dephasing_start}
\end{aligned}
\end{equation}

The first term is trivially:
\begin{equation}
\expval{{b_{\rm in}}^\dagger(\omega)b_{\rm in}(\omega)} = \abs{f(\omega)}^2 ,\label{eq:part1}
\end{equation}
while the second term is
\begin{equation}
\begin{aligned}
    \sqrt{\gamma}\expval{b_{\rm in}^\dagger(\omega)\sigma(\omega)} &= -\sqrt{\gamma}f^*(\omega) \bra{0,g} \sigma(\omega) \ket{1,g} \\ 
    &= -\sqrt{\gamma}f^*(\omega) c(\omega),  
\end{aligned}
\end{equation}
 where we defined $c(\omega) = \bra{0,g} \sigma(\omega) \ket{1,g}$, which we find by writing the equation of motion:
 \begin{equation}
\begin{aligned}
    \dot c(t)
&=
i \bra{0,g} \comm{H}{\sigma(t)} \ket{1,g}
+
\gamma_\phi
\bra{0,g}
D_{\sigma^\dagger\sigma}^\dagger[\sigma(t)]
\ket{1,g}.\\
&= 
(i\delta-\Gamma)c(t)
-
\sqrt{\gamma}f(t).
\label{eq:c_eom_dephasing}    
\end{aligned}
\end{equation}
with $\Gamma = (\gamma + \gamma_\phi)/2$,  $\delta = \omega_0 - \omega_p$ being the detuning between the photon pulse and emitter transition frequency $\omega_0$, and $D^\dagger_A(\rho) = A^\dagger \rho A - 1/2(A^\dagger A \rho + \rho A^\dagger A)$ is the conjugate Lindblad operator, which occurs from reordering the trace. 

Fourier transforming Eq.~\eqref{eq:c_eom_dephasing}, we get:
\begin{equation}
c(\omega)
=
-
\frac{\sqrt{\gamma}f(\omega)}
{\Gamma+i(\omega-\delta)} .
\label{eq:c_omega_dephasing}
\end{equation}
Thus, we have:
\begin{equation}\sqrt{\gamma}\expval{b_{in}^\dagger(\omega)\sigma(\omega)} = \frac{\gamma \abs{f(\omega)}^2}
{\Gamma+i(\omega-\delta)}. \label{eq:part2}
\end{equation}
Similarly, $\expval{\sigma^\dagger(\omega) b_{\rm in}(\omega) } = -\sqrt{\gamma} f(\omega) c^*(\omega)$
and thus:
\begin{equation}
    \sqrt{\gamma} \expval{\sigma^\dagger(\omega) b_{\rm in}(\omega) } = \frac{\gamma \abs{f(\omega)}^2}
{\Gamma - i(\omega-\delta)} .\label{eq:part3}
\end{equation}

The final term $\expval{\sigma^\dagger(\omega)\sigma(\omega)}$ is more cumbersome and requires additional detail when pure dephasing is introduced. We write:
\begin{equation}
\begin{split}
\left\langle
\sigma^\dagger(\omega)\sigma(\omega)
\right\rangle
&=
\int dt ds 
e^{i\omega t}e^{-i\omega s}
\left\langle
\sigma^\dagger(t)\sigma(s)
\right\rangle .
\end{split}
\label{eq:sigma_sigma_frequency_twotime}
\end{equation}
Therefore, we need to compute the two-time correlation function:
\begin{equation}
G(t,s)
=
\left\langle
1,g
\right|
\sigma^\dagger(t)\sigma(s)
\left|
1,g
\right\rangle.
\end{equation}

The equation of motion for \(\sigma(s)\) is
\begin{equation}
\frac{d}{ds}\sigma(s)
=
(i\delta-\Gamma)\sigma(s)
-
\sqrt{\gamma} b_{\rm in}(s)\sigma_z(s).
\end{equation}
Multiplying by $\sigma^\dagger(t)$ from the left and taking the one-photon expectation value gives
\begin{equation}
\begin{split}
\frac{\partial}{\partial s}G(t,s)
&=
(i\delta-\Gamma)G(t,s)
-
\sqrt{\gamma}
\left\langle
\sigma^\dagger(t)b_{\rm in}(s)\sigma_z(s)
\right\rangle .
\end{split}
\end{equation}

For a one-photon input pulse, the source term evaluates to
\begin{equation}
-\sqrt{\gamma}
\left\langle
\sigma^\dagger(t)b_{\rm in}(s)\sigma_z(s)
\right\rangle
=
-\sqrt{\gamma} f(s)c^*(t).
\end{equation}
Therefore, the two-time correlation function obeys
\begin{equation}
\frac{\partial}{\partial s}G(t,s)
=
(i\delta-\Gamma)G(t,s)
-
\sqrt{\gamma} f(s)c^*(t).
\label{eq:G_eom_dephasing}
\end{equation}

The initial condition is
\begin{equation}
G(t,t)
=
\left\langle
\sigma^\dagger(t)\sigma(t)
\right\rangle
=
P(t),
\end{equation}
where
\begin{equation}
P(t)
=
\bra{1,g}
\sigma^\dagger(t)\sigma(t)
\ket{1,g}.
\end{equation}
Solving Eq.~\eqref{eq:G_eom_dephasing} for \(s\geq t\) gives
\begin{equation}
\begin{aligned}
G(t,s)
&=
e^{(i\delta-\Gamma)(s-t)}P(t)
\\
&\quad
-
\sqrt{\gamma} c^*(t)
\int_t^s du\,
e^{(i\delta-\Gamma)(s-u)}f(u).
\end{aligned}
\label{eq:G_solution_raw}
\end{equation}

On the other hand, the solution of Eq.~\eqref{eq:c_eom_dephasing} between \(t\) and \(s\) is
\begin{equation}
\begin{aligned}
c(s)
&=
e^{(i\delta-\Gamma)(s-t)}c(t)
\\
&\quad
-
\sqrt{\gamma}
\int_t^s du\,
e^{(i\delta-\Gamma)(s-u)}f(u).
\end{aligned}
\label{eq:c_s_from_t}
\end{equation}
Multiplying Eq.~\eqref{eq:c_s_from_t} by \(c^*(t)\) gives
\begin{equation}
\begin{aligned}
c^*(t)c(s)
&=
e^{(i\delta-\Gamma)(s-t)}
\abs{c(t)}^2
\\
&\quad
-
\sqrt{\gamma} c^*(t)
\int_t^s du\,
e^{(i\delta-\Gamma)(s-u)}f(u).
\end{aligned}
\end{equation}
Subtracting this expression from Eq.~\eqref{eq:G_solution_raw}, the integral terms cancel, yielding
\begin{equation}
G(t,s)-c^*(t)c(s)
=
e^{(i\delta-\Gamma)(s-t)}
\left[
P(t)-\abs{c(t)}^2
\right].
\end{equation}

Thus, the two-time correlation function can be decomposed as
\begin{equation}
G(t,s)
=
c^*(t)c(s)
+
Q(t)e^{(i\delta-\Gamma)(s-t)},
\label{eq:G_split_dephasing}
\end{equation}
where
\begin{equation}
Q(t)
=
P(t)-\abs{c(t)}^2.
\label{eq:Q_def}
\end{equation}

The population obeys
\begin{equation}
\dot P(t)
=
-\gamma P(t)
-
\sqrt{\gamma}
\left[
f^*(t)c(t)+f(t)c^*(t)
\right].
\label{eq:P_eom_dephasing}
\end{equation}
Meanwhile,
\begin{equation}
\frac{d}{dt}\abs{c(t)}^2
=
-2\Gamma \abs{c(t)}^2
-
\sqrt{\gamma}
\left[
f^*(t)c(t)+f(t)c^*(t)
\right].
\end{equation}
Here we already see that, if \(\gamma_\phi=0\), then \(P(t)=\abs{c(t)}^2\), since the two quantities obey the same equation of motion. When \(\gamma_\phi\neq0\), however, this is no longer the case, and \(Q(t)\neq0\). This term corresponds to incoherent emission from the emitter. 

Using \(Q(t)=P(t)-\abs{c(t)}^2\), we find
\begin{equation}
\begin{aligned}
\dot Q(t)
&=
\dot P(t)
-
\frac{d}{dt}\abs{c(t)}^2
\\
&=
-\gamma P(t)
+
2\Gamma \abs{c(t)}^2
\\
&=
-\gamma Q(t)
+
(2\Gamma-\gamma)\abs{c(t)}^2 \\
&= -\gamma Q(t)
+
\gamma_\phi \abs{c(t)}^2. \label{eq:Q_eom}
\end{aligned}
\end{equation}
Thus, \(Q(t)\) is generated only by pure dephasing. If \(\gamma_\phi=0\), then \(Q(t)=0\) for the one-photon problem. In this case, \(G(t,s)=c^*(t)c(s)\), and therefore
\(\left\langle\sigma^\dagger(\omega)\sigma(\omega)\right\rangle
=
\abs{c(\omega)}^2\). For \(\gamma_\phi=0\), this recovers the result of Ref.~\cite{ArranzRegidor2025TheorySpectra}. The coherent contribution also agrees with Ref.~\cite{Ramos2018}, where only the coherently scattered part is given for \(\gamma_\phi\neq0\).

We now insert Eq.~\eqref{eq:G_split_dephasing} into the frequency-domain atomic spectrum. Writing \(s=t+\tau\), we have
\begin{equation}
\begin{aligned}
\left\langle
\sigma^\dagger(\omega)\sigma(\omega)
\right\rangle
&=
2\,\mathrm{Re}
\int_{-\infty}^{\infty} dt
\int_0^\infty d\tau\,
e^{-i\omega\tau}
G(t,t+\tau).
\end{aligned}
\label{eq:sigma_spectrum_tau}
\end{equation}
Using Eq.~\eqref{eq:G_split_dephasing}, this becomes
\begin{equation}
\begin{aligned}
\left\langle
\sigma^\dagger(\omega)\sigma(\omega)
\right\rangle
&=
2\,\mathrm{Re}
\int dt
\int_0^\infty d\tau\,
e^{-i\omega\tau}
c^*(t)c(t+\tau)
\\
&\quad
+
2\,\mathrm{Re}
\int dt
\int_0^\infty d\tau\,
e^{-i\omega\tau}
Q(t)e^{(i\delta-\Gamma)\tau}.
\end{aligned}
\end{equation}

The first term is the coherent part, as also mentioned above, it gives
\begin{equation}
\left\langle
\sigma^\dagger(\omega)\sigma(\omega)
\right\rangle_{\rm coh}
=
\abs{c(\omega)}^2.
\end{equation}
Using Eq.~\eqref{eq:c_omega_dephasing}, we have
\begin{equation}
\abs{c(\omega)}^2
=
\frac{\gamma \abs{f(\omega)}^2}
{(\omega-\delta)^2+\Gamma^2}.
\end{equation}
Therefore,
\begin{equation}
\gamma
\left\langle
\sigma^\dagger(\omega)\sigma(\omega)
\right\rangle_{\rm coh}
=
\frac{\gamma^2 \abs{f(\omega)}^2}
{(\omega-\delta)^2+\Gamma^2}.
\label{eq:coherent_atomic_spectrum}
\end{equation}
The second term is the incoherent part. It is
\begin{equation}
\begin{aligned}
\left\langle
\sigma^\dagger(\omega)\sigma(\omega)
\right\rangle_{\rm inc}
&=
2\,\mathrm{Re}
\int_{-\infty}^\infty dt\, Q(t)
\int_0^\infty d\tau\,
e^{-i\omega\tau}
e^{(i\delta-\Gamma)\tau}
\\
&=
2\,\mathrm{Re}
\left[
\frac{1}
{\Gamma+i(\omega-\delta)}
\right]
\int_{-\infty}^\infty dt\, Q(t)
\\
&=
\frac{2\Gamma}
{(\omega-\delta)^2+\Gamma^2}
\int_{-\infty}^\infty dt\, Q(t).
\end{aligned}
\label{eq:incoherent_sigma_spectrum}
\end{equation}
Thus,
\begin{equation}
\gamma 
\left\langle
\sigma^\dagger(\omega)\sigma(\omega)
\right\rangle_{\rm inc}
=
\frac{2\gamma\Gamma}
{(\omega-\delta)^2+\Gamma^2}
\int_{-\infty}^\infty dt\, Q(t).
\label{eq:incoherent_atomic_spectrum}
\end{equation}

We can express the integrated incoherent population in terms of \(c(t)\). Integrating Eq.~\eqref{eq:Q_eom} over time, and assuming that \(Q(t)\) vanishes at the temporal boundaries, gives
\begin{equation}
0
=
-\gamma \int_{-\infty}^\infty dt\, Q(t)
+
\gamma_\phi
\int_{-\infty}^\infty dt\, \abs{c(t)}^2.
\end{equation}
Hence,
\begin{equation}
\int_{-\infty}^\infty dt\, Q(t)
=
\frac{\gamma_\phi}{\gamma}
\int_{-\infty}^\infty dt\, \abs{c(t)}^2.
\label{eq:int_Q}
\end{equation}

Using Parseval's theorem,
\begin{equation}
\int_{-\infty}^\infty dt\, \abs{c(t)}^2
=
\int_{-\infty}^\infty \frac{d\nu}{2\pi}
\abs{c(\nu)}^2
=
\int_{-\infty}^\infty \frac{d\nu}{2\pi}
\frac{\gamma \abs{f(\nu)}^2}
{(\nu-\delta)^2+\Gamma^2}.
\end{equation}
Therefore,
\begin{equation}
\int dt\, Q(t)
=
\gamma_\phi
\int \frac{d\nu}{2\pi}
\frac{\abs{f(\nu)}^2}
{(\nu-\delta)^2+\Gamma^2}.
\label{eq:int_Q_final}
\end{equation}

Substituting Eq.~\eqref{eq:int_Q_final} into Eq.~\eqref{eq:incoherent_atomic_spectrum}, we obtain
\begin{equation}
\gamma
\left\langle
\sigma^\dagger(\omega)\sigma(\omega)
\right\rangle_{\rm inc}
=
\frac{2\gamma\gamma_\phi\Gamma}
{(\omega-\delta)^2+\Gamma^2}
\int \frac{d\nu}{2\pi}
\frac{\abs{f(\nu)}^2}
{(\nu-\delta)^2+\Gamma^2}.
\label{eq:S_inc_final}
\end{equation}
Combining the coherent and incoherent atomic contributions gives
\begin{equation}
\begin{aligned}
\gamma
\left\langle
\sigma^\dagger(\omega)\sigma(\omega)
\right\rangle
&=
\frac{\gamma^2 \abs{f(\omega)}^2}
{(\omega-\delta)^2+\Gamma^2}
\\
&\quad
+
\frac{2\gamma\gamma_\phi\Gamma}
{(\omega-\delta)^2+\Gamma^2}
\int \frac{d\nu}{2\pi}
\frac{\abs{f(\nu)}^2}
{(\nu-\delta)^2+\Gamma^2}.
\end{aligned}
\label{eq:atomic_total_dephasing}
\end{equation}

Finally, inserting Eq.~\eqref{eq:atomic_total_dephasing} into Eq.~\eqref{eq:stationary_spectrum_dephasing_start}, together with Eqs.~\eqref{eq:part1},\eqref{eq:part2}, and \eqref{eq:part3}, gives
\begin{equation}
\begin{aligned}
S(\omega)
&=
\left[
1
-
\frac{2\gamma\Gamma}
{(\omega-\delta)^2+\Gamma^2}
+
\frac{\gamma^2}
{(\omega-\delta)^2+\Gamma^2}
\right]
\abs{f(\omega)}^2
\\
&\quad
+
\frac{2\gamma\gamma_\phi\Gamma}
{(\omega-\delta)^2+\Gamma^2}
\int \frac{d\nu}{2\pi}
\frac{\abs{f(\nu)}^2}
{(\nu-\delta)^2+\Gamma^2}.
\end{aligned}
\label{eq:stationary_chiral_dephasing}
\end{equation}

Since
\begin{equation}
t(\omega)
=
1-\frac{\gamma}{\Gamma+i(\omega-\delta)}
=
\frac{\Gamma-\gamma+i(\omega-\delta)}
{\Gamma+i(\omega-\delta)},
\end{equation}
we have
\begin{equation}
\begin{aligned}
\abs{t(\omega)}^2
&=
\frac{
(\Gamma-\gamma)^2+(\omega-\delta)^2
}
{
\Gamma^2+(\omega-\delta)^2
}
\\
&=
1
-
\frac{2\gamma\Gamma-\gamma^2}
{(\omega-\delta)^2+\Gamma^2}.
\end{aligned}
\end{equation}

Therefore, the stationary spectrum can be written as
\begin{equation}
\begin{aligned}
S(\omega)
&=
\abs{t(\omega)}^2\abs{f(\omega)}^2
\\
&\quad
+
\frac{2\gamma\gamma_\phi\Gamma}
{(\omega-\delta)^2+\Gamma^2}
\int \frac{d\nu}{2\pi}
\frac{\abs{f(\nu)}^2}
{(\nu-\delta)^2+\Gamma^2}.
\end{aligned}
\label{eq:stationary_chiral_dephasing_compact}
\end{equation}
which is the expression given in Eq.~\eqref{eq:s_total} of the main text. When \(\gamma_\phi=0\), we have \(\Gamma=\gamma/2\), while the incoherent term vanishes, and Eq.~\eqref{eq:stationary_chiral_dephasing_compact} reduces to
\begin{equation}
S_R(\omega)
=
\abs{f(\omega)}^2,
\end{equation}
which is the stationary chiral result without pure dephasing.

\end{document}